\documentclass[manuscript]{acmart}

\AtBeginDocument{%
  \providecommand\BibTeX{{%
    \normalfont B\kern-0.5em{\scshape i\kern-0.25em b}\kern-0.8em\TeX}}}

\setcopyright{acmcopyright}
\copyrightyear{2022}
\acmYear{2022}
\acmDOI{}

\usepackage{tabularx}
\usepackage{subcaption}
\usepackage{multirow} 
\usepackage{array} 
\usepackage{comment} 
\usepackage{amsmath}
\usepackage[]{algorithm2e}
\usepackage{xcolor}

\setcopyright{none}

\begin{document}

\title{Scale-out Systolic Arrays}

\author{Ahmet Caner Y\"{u}z\"{u}g\"{u}ler}
\email{ahmet.yuzuguler@epfl.ch}
\orcid{ 0000-0001-7809-9897 }
\affiliation{%
  \institution{EPFL}
  \country{Switzerland}
}

\author{Canberk S\"{o}nmez}
\email{canberk.sonmez@epfl.ch}
\affiliation{%
  \institution{EPFL}
  \country{Switzerland}
}

\author{Mario Drumond}
\email{mario.drumond@pm.me}
\affiliation{%
  \institution{CodeDepot}
  \country{Switzerland}
}

\author{Yunho Oh}
\email{yunho.oh@skku.edu}
\affiliation{%
  \institution{SungKyunKwan University}
  \country{South Korea}
}

\author{Babak Falsafi}
\email{babak.falsafi@epfl.ch}
\affiliation{%
  \institution{EPFL}
  \country{Switzerland}
}

\author{Pascal Frossard}
\email{pascal.frossard@epfl.ch}
\affiliation{%
  \institution{EPFL}
  \country{Switzerland}
}

\renewcommand{\shortauthors}{Y\"{u}z\"{u}g\"{u}ler, et al.}

\newcommand{\multirowcell}[2]{\begin{tabular}[c]{@{}c@{}} #1\\#2\end{tabular}}

\newcommand{\multrowC}[1]{\begin{tabular}[c]{@{}c@{}} #1 \end{tabular}}
\newcommand{\multrowR}[1]{\begin{tabular}[c]{@{}r@{}} #1 \end{tabular}}
\newcommand{\multrowL}[1]{\begin{tabular}[c]{@{}l@{}} #1 \end{tabular}}
\newcommand{\unit}[1]{\lbrack#1\rbrack}

\begin{abstract}
Multi-pod systolic arrays are emerging as the architecture of choice in DNN inference accelerators. Despite their potential, designing multi-pod systolic arrays to maximize effective throughput/Watt---i.e., throughput/Watt adjusted when accounting for array utilization---poses a unique set of challenges. In this work, we study three key pillars in multi-pod systolic array designs, namely array granularity, interconnect, and tiling. We identify optimal array granularity across workloads and show that state-of-the-art commercial accelerators use suboptimal array sizes for single-tenancy workloads. We, then evaluate the bandwidth/latency trade-offs in interconnects and show that Butterfly networks offer a scalable topology for accelerators with a large number of pods. Finally, we introduce a novel data tiling scheme with custom partition size to maximize utilization in optimally sized pods. We propose \emph{Scale-out Systolic Arrays}, a multi-pod inference accelerator for both single- and multi-tenancy based on these three pillars. We show that SOSA exhibits scaling of up to 600 TeraOps/s in effective throughput for state-of-the-art DNN inference workloads, and outperforms state-of-the-art multi-pod accelerators by a factor of $1.5 \times$.
\end{abstract}



\maketitle

\def\SynthAbsolutePowerSRAM{ 132710.40 }
\def\SynthAbsolutePowerPPFX{ 1609.74 }
\def\SynthAbsolutePowerInterconnect{ 43632.41 }
\def\SynthAbsolutePowerPodArray{ 109047.04 }
\def\SynthAbsolutePowerPodJobQueue{ 875.52 }
\def\SynthAbsolutePowerPodActBuffer{ 193.54 }
\def\SynthAbsolutePowerPodConvBuffer{ 542.21 }
\def\SynthAbsolutePowerPodInputPsumBuffer{ 274.69 }
\def\SynthAbsolutePowerPodOutputPsumBuffer{ 267.52 }
\def\SynthAbsolutePowerPodOthers{ 548.61 }
\def\SynthPercentagePowerSRAM{ 45.81 }
\def\SynthPercentagePowerPPFX{ 0.56 }
\def\SynthPercentagePowerInterconnect{ 15.06 }
\def\SynthPercentagePowerPodArray{ 37.64 }
\def\SynthPercentagePowerPodJobQueue{ 0.30 }
\def\SynthPercentagePowerPodActBuffer{ 0.07 }
\def\SynthPercentagePowerPodConvBuffer{ 0.19 }
\def\SynthPercentagePowerPodInputPsumBuffer{ 0.09 }
\def\SynthPercentagePowerPodOutputPsumBuffer{ 0.09 }
\def\SynthPercentagePowerPodOthers{ 0.19 }
\def\SynthAbsoluteAreaSRAM{ 402822912.00 }
\def\SynthAbsoluteAreaPPFX{ 1326683.74 }
\def\SynthAbsoluteAreaInterconnect{ 22346081.32 }
\def\SynthAbsoluteAreaPodArray{ 105635405.21 }
\def\SynthAbsoluteAreaPodJobQueue{ 985654.66 }
\def\SynthAbsoluteAreaPodActBuffer{ 71663.62 }
\def\SynthAbsoluteAreaPodConvBuffer{ 296358.91 }
\def\SynthAbsoluteAreaPodInputPsumBuffer{ 143327.23 }
\def\SynthAbsoluteAreaPodOutputPsumBuffer{ 143327.23 }
\def\SynthAbsoluteAreaPodOthers{ 718440.19 }
\def\SynthPercentageAreaSRAM{ 75.37 }
\def\SynthPercentageAreaPPFX{ 0.25 }
\def\SynthPercentageAreaInterconnect{ 4.18 }
\def\SynthPercentageAreaPodArray{ 19.76 }
\def\SynthPercentageAreaPodJobQueue{ 0.18 }
\def\SynthPercentageAreaPodActBuffer{ 0.01 }
\def\SynthPercentageAreaPodConvBuffer{ 0.06 }
\def\SynthPercentageAreaPodInputPsumBuffer{ 0.03 }
\def\SynthPercentageAreaPodOutputPsumBuffer{ 0.03 }
\def\SynthPercentageAreaPodOthers{ 0.13 }
\def\SystolicArrayPowerRelativePod{ 97.58 }
\def\SystolicArrayAreaRelativePod{ 97.82 }

\section{Introduction}
\label{sec:intro}
Deep Neural Networks (DNN) are widely employed in various domains from computer vision to natural language processing. With the slowdown in silicon scaling and the increase in demand for throughput with a tight tail latency, designers have resorted to custom hardware accelerators for DNN  workloads \cite{sze17,Han2016,Kwon19,farabet2011,sharma2016, chen19,fowers18,ovtcharov15}. While many prior work have adopted spatial architectures that implement certain dataflows (e.g., Eyeriss~\cite{chen2016}, MAERI~\cite{kwon18}, SIGMA~\cite{Qin20}), systolic arrays have emerged as the architecture of choice in commercial silicon products (e.g., Google's TPU~\cite{jouppi2017}, Tesla's FSD chip~\cite{Bannon19}) thanks to their exceptional power efficiency.

Since their inception, systolic arrays have dramatically evolved to achieve remarkable peak throughput with higher levels of silicon integration, power envelopes, and memory bandwidth. Unfortunately, translating peak throughput to higher utilization has remained challenging for designers because of mismatches between workload requirements and array size. Google’s TPU v1 and v4 reported utilizations of about 22\%~\cite{jouppi2017} and 33\%~\cite{Jouppi2021}, respectively, while processing DNN models due their large systolic array-based matrix multiplication unit.

To improve the total cost of ownership and support multi-tenancy, recent accelerators---such as TPU v3 and v4---incorporate a few coarse-grain systolic-array \emph{pods} connected through shared memory on a single die~\cite{Jouppi2021,baek2020}. While these \emph{multi-pod} accelerators achieve much better utilization over their monolithic counterparts with multi-tenancy, variability in array size requirements in workloads remains a fundamental limitation to utilization in a few coarse-grain pods. In contrast, multi-pod designs with minimally sized arrays~\cite{kung2019} target maximum utilization. Unfortunately, these designs compromise the inference accelerator's power efficiency by over-provisioning overall on-chip memory~\cite{Drumond2021} (e.g., 8x8 arrays incur $5-10\times$ more memory accesses than $128\times128$ arrays). Therefore, even at high utilization, such multi-pod designs achieve inferior throughput/Watt relative to designs with coarse-grain pods~\cite{Samajdar18}.

In this paper, we make the observation that optimally sizing systolic arrays in multi-pod accelerators is key to achieving optimal \emph{effective} throughput/Watt---i.e., the throughput/Watt adjusted when accounting for array utilization---in single workloads. Much like multi-threaded CPUs targeting high single-threaded performance, multi-pod accelerators with optimally sized pods seamlessly support efficient multi-tenancy in workloads.

A multi-pod inference accelerator with a large number of optimally sized pods requires a scalable interconnect with favorable bandwidth and latency characteristics. Much prior work has focused on the use of interconnects in inference accelerators. While many advocate Mesh~\cite{Chen17, Chen14, gao19, shao19} or H-tree \cite{kung2019, Stevens19}, these topologies lack sufficient bisection bandwidth to support a large number of pods.  Others advocate Benes~\cite{Qin20}, which requires a long round-trip latency on requests and may adversely impact the overall execution time. As such, interconnecting a large number of pods remains an open research problem. Similarly, an accelerator with a large number of optimally sized pods also creates an opportunity to customize partitioning for input activations. Prior work using multi-pod accelerators either does not partition the input \cite{baek2020} or uses sub-optimal partitions \cite{Choi20}. We make the observation that an input partition size that is optimal for a given array granularity exposes the available data parallelism within the DNN layers, thereby maximizing the available number of tiles that can be extracted from a single workload.

This paper proposes \emph{Scale-out Systolic Arrays (SOSA)}, a multi-pod architecture supporting both single- and multi-tenancy workloads. SOSA incorporates optimally sized systolic arrays for a spectrum of state-of-the-art inference workloads from computer vision to NLP. SOSA relies on a Butterfly network as a scalable fabric to interconnect systolic arrays and memory banks with a high bisection and low latency, and employs a novel data tiling scheme that maximizes the available number of tiles for a given array size. We use in-house simulators with timing models and synthesis tools for area and power to show that SOSA outperforms both coarse- and fine-grain multi-pod inference accelerators while achieving strong scalability.

In this paper, we make the following contributions:

\begin{itemize}
    \item
     Our design space exploration across a wide spectrum of workloads indicates that SOSA with $32 \times 32$ pods achieves $1.5\times$ higher effective throughput/Watt than $128 \times 128$ pods used in state-of-art commercial accelerators (e.g., TPUs~\cite{tpu2017}) for single workloads.
    
    \item
    Our baseline SOSA with a Butterfly network achieves $2\times$ higher effective throughput than one with a Benes network and $2.3\times$ less power than one with a crossbar.
    
    \item
    Customizing the tiling strategy in multi-pod accelerators improves utilization by up to $5\times$ over a system with no partitioning for activations.
    
    \item We show that the proposed design can achieve strong scaling up to 600 TeraOps/s at a TDP of 400 Watts for computationally-intensive DNN models such as Resnet.
    
    \item 
    SOSA increases effective throughput by $1.44\times$ with multi-tenancy over single-tenancy.
    
\end{itemize}

For reproducibility, we open-source our cycle-accurate DNN accelerator simulator.\footnote{\url{https://github.com/yuezuegu/sosa-compiler}} To the best of our knowledge, this is the first open-source simulator for multi-pod systolic array accelerators.

\section{Why Scale-Out Systolic Arrays?} \label{sec:whysosa}

\begin{figure}[t]
    \centering
    \includegraphics[width=0.5\linewidth]{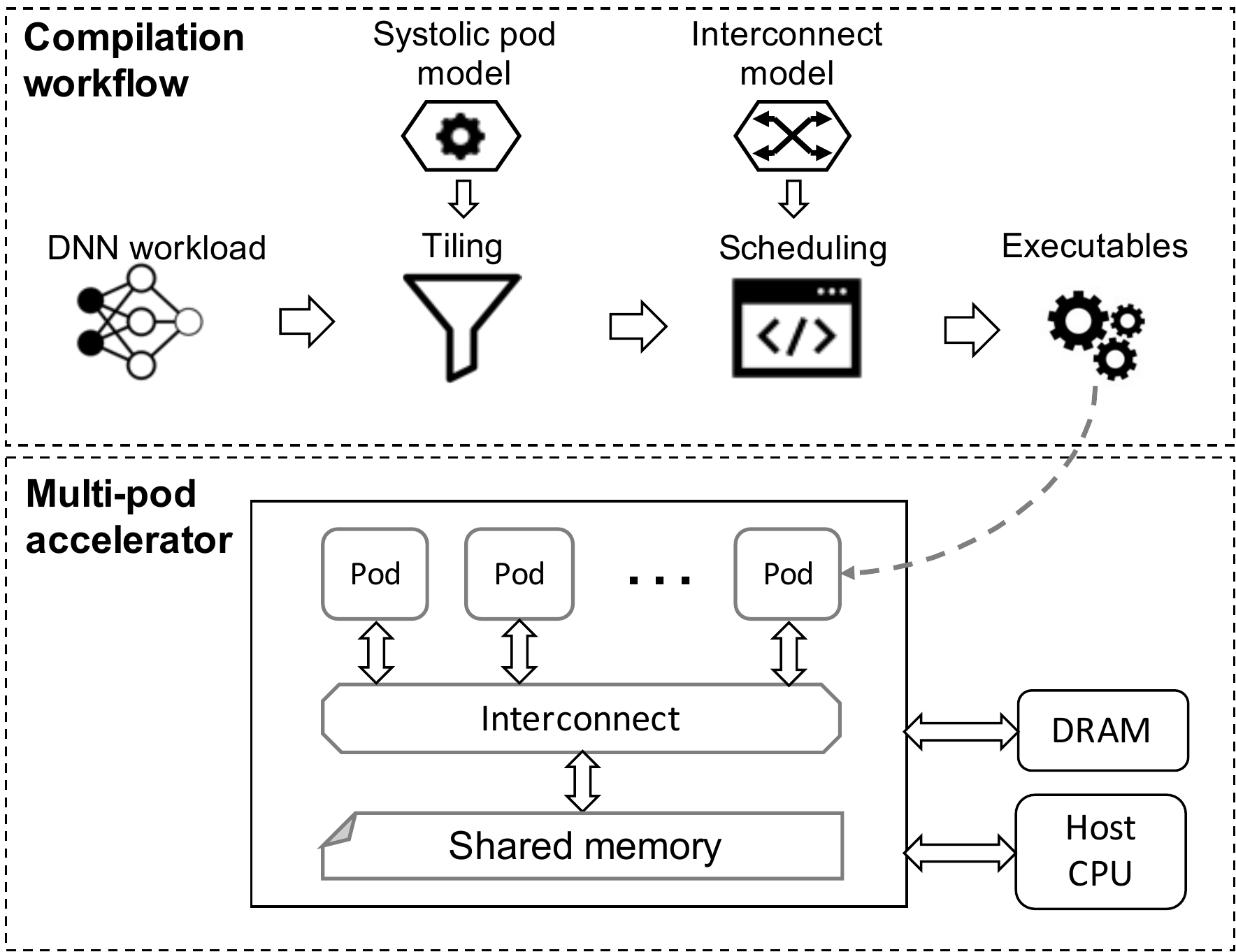}
	\caption{General overview of a multi-pod systolic array.}
    \label{fig:overview}
\end{figure}

A myriad of prior work have proposed dataflow optimizations such as row-stationary scheduling~\cite{chen2016}, flexible interconnection~\cite{lu17, kwon18, Qin20}, and novel data partitioning and scheduling schemes~\cite{Kwon19, gao19}. These optimizations have two common microarchitectural requirements. First, each processing element must have a scratchpad memory to reuse data temporally. Second, interconnection between processing elements must be reconfigurable to support efficient data mapping and scheduling schemes. As a result of these two requirements, majority of power consumption per MAC operation is spent on memory access and data movement~\cite{chen2016, Kwon19, gao19}, which places an upper bound on the power efficiency of such accelerators.

Systolic arrays overcome the limitations of dataflow architectures by adopting simplistic processing element microarchitectures (i.e., without large register files or scratchpad memory) and implementing static interconnection between processing elements. As such, systolic arrays achieve higher power efficiency and peak throughput compared to dataflow architectures. Moreover, recent proposals to couple multiple systolic arrays in a single die (i.e., \emph{multi-pod} designs~\cite{kung2019,baek2020,Jouppi2021}) allows benefiting data and task-level parallelism, further improving the gain from provisioned silicon.

Figure \ref{fig:overview} depicts the overall diagram of a multi-pod inference accelerator\cite{kung2019, tpu2017, baek2020}, where each \emph{pod} includes a systolic array and the necessary glue interface to an interconnect connecting the pods together to memory and the peripherals. Like recent custom inference accelerators (e.g., Graphcore\cite{graphcore2017}, Brainwave\cite{fowers18}), models and data are stored in on-chip memory with an interface to off-chip DRAM and a host CPU. To map workloads to multiple systolic arrays, DNN layers are first partitioned into tile operations of sizes that match a pod's array dimensions. Using information about connectivity and latency about the interconnect, a static scheduler optimizes the mapping of tile operations to pods to maximize parallelism and throughput. 

\begin{figure}[t]
    \centering
    \includegraphics[width=0.8\linewidth]{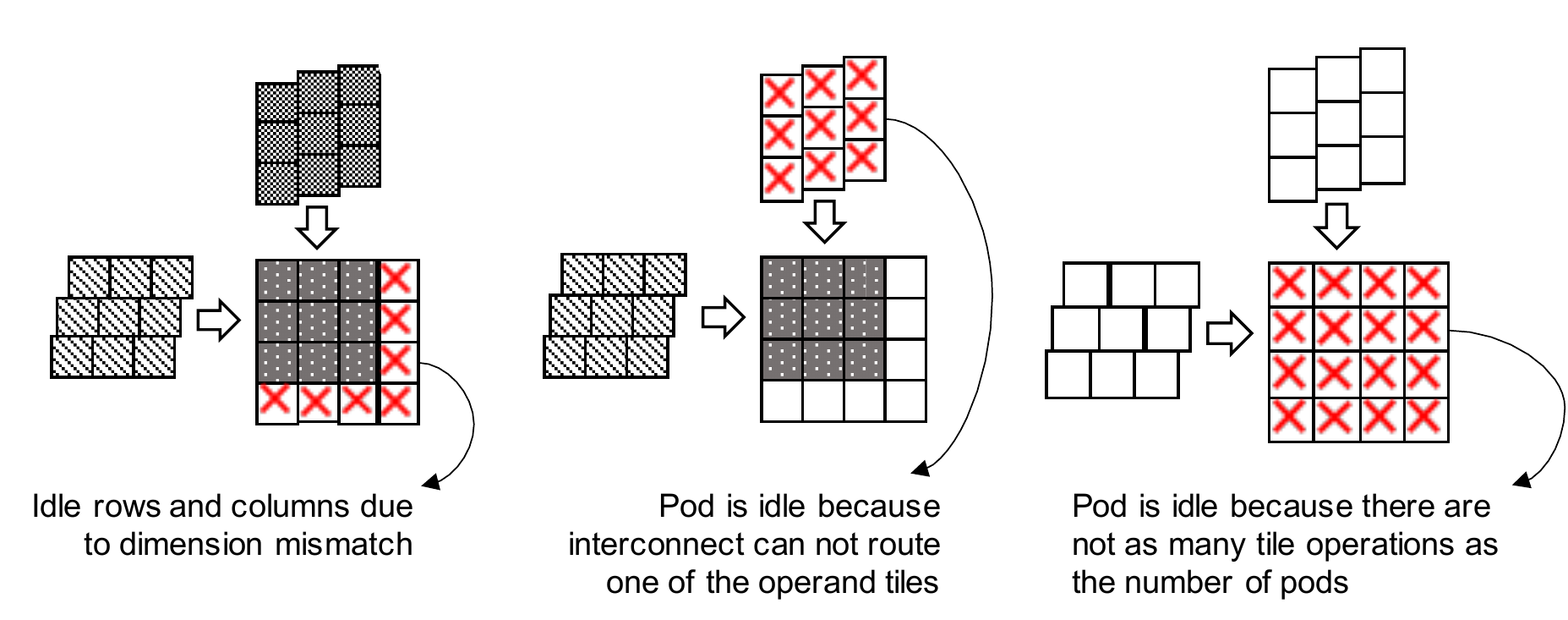}
	\caption{Three main factors of underutilization in systolic arrays. }
    \label{fig:utilization}
\end{figure}

A multi-pod accelerator's effective throughput is a function of the overall utilization of processing elements both within and across pods. Therefore, maintaining a high utilization not only maximizes the gain from provisioned silicon resources but also improves the overall performance of an accelerator. Figure \ref{fig:utilization} illustrates three main causes of the underutilization in a multi-pod accelerator: (1) dimension mismatch \cite{Samajdar18} between array and workload resulting in underutilization within a pod, (2) poor connectivity resulting in underutilization across pods, and (3) sub-optimal tiling resulting in underutilization both within and across pods.


Utilization within a pod highly depends on the pod's systolic array granularity and the layer dimensions of a DNN workload. Prior multi-pod accelerators \cite{jouppi2017, tpu2017, baek2020} opt for larger array dimensions to reduce access to on-chip memory, provisioning higher power for processing elements. Unfortunately, larger arrays also increase the likelihood that the workload's layer dimensions are smaller than the number of array's rows or columns, resulting in idle processing elements and wasted throughput/Watt. In contrast, minimizing the array dimensions per pod reduces the mismatch between the workloads and the array, resulting in improved utilization. Smaller systolic arrays, however, increase the power required for on-chip memory access, undermining the overall throughput/Watt. In this paper, we show that the optimal array size for DNN workloads is an order of magnitude smaller than that widely adopted by academia and industry.  

The interconnect also plays a key role in utilization and effective throughput in multi-pod accelerators \cite{kung2019}. To achieve a scalable multi-pod accelerator architecture with high utilization, we identify four design requirements for the interconnect. First, the interconnect should provide high bisection bandwidth among pods and on-chip memory to enable a continuous flow of operands and partial results. Second, the interconnect should support high combinatorial power to allow for distributing operands and partial results with minimal contention and delay. Third, the interconnect should have a short round-trip latency to help hide the data movement latency with computation. Finally, because optimally sized arrays can lead to a relatively large number (e.g., hundreds) of connected pods, silicon provisioning for the interconnect should also scale well with the increasing number of pods. 


Finally, the tile size for each pod fundamentally impacts utilization in multi-pod accelerators and should be tuned with care. We observe that conventional approaches to tiling for systolic arrays \cite{baek2020, Choi20} fall well short of generating a sufficient number of tile operations to populate a large number of pods resulting in idle arrays during execution. On the one hand, choosing large tiles limits the overall number of tile operations and results in idle pods. On the other hand, choosing a small tiling size may introduce underutilization within pods due to internal buffering times.

\section{Key Pillars of Multi-pod Accelerator Design} \label{sec:optimal-size}


In Section \ref{sec:whysosa}, we argued that the optimal array size of systolic arrays is smaller than those in conventional DNN accelerators, which gives rise to multi-pod architectures. In this section, we introduce our approach regarding the key pillars of energy-efficient multi-pod DNN accelerators. First, we show that the optimal systolic array size is a function of data dimensions in target DNN workloads and perform a design space exploration of the optimal array sizes for various scenarios. Second, we introduce an analysis of various interconnection networks and our decision for the multi-pod DNN accelerators.

\subsection{Optimal Systolic Array Size} \label{sec:optimal-size}
DNN models typically comprise a number of layers of various types, such as convolutional, fully-connected, and attention, whose computation can be expressed as general matrix multiplication (GEMM) operations. Without loss of generality, we assume that GEMM operations are in the form of $XW+P_{in} = P_{out}$, where $X$ and $W$ denote DNN activations and weights; $P_{in}$ and $P_{out}$ denote input and output partial sums, respectively.

\begin{figure}[t]
    \centering
    \includegraphics[width=0.6\linewidth]{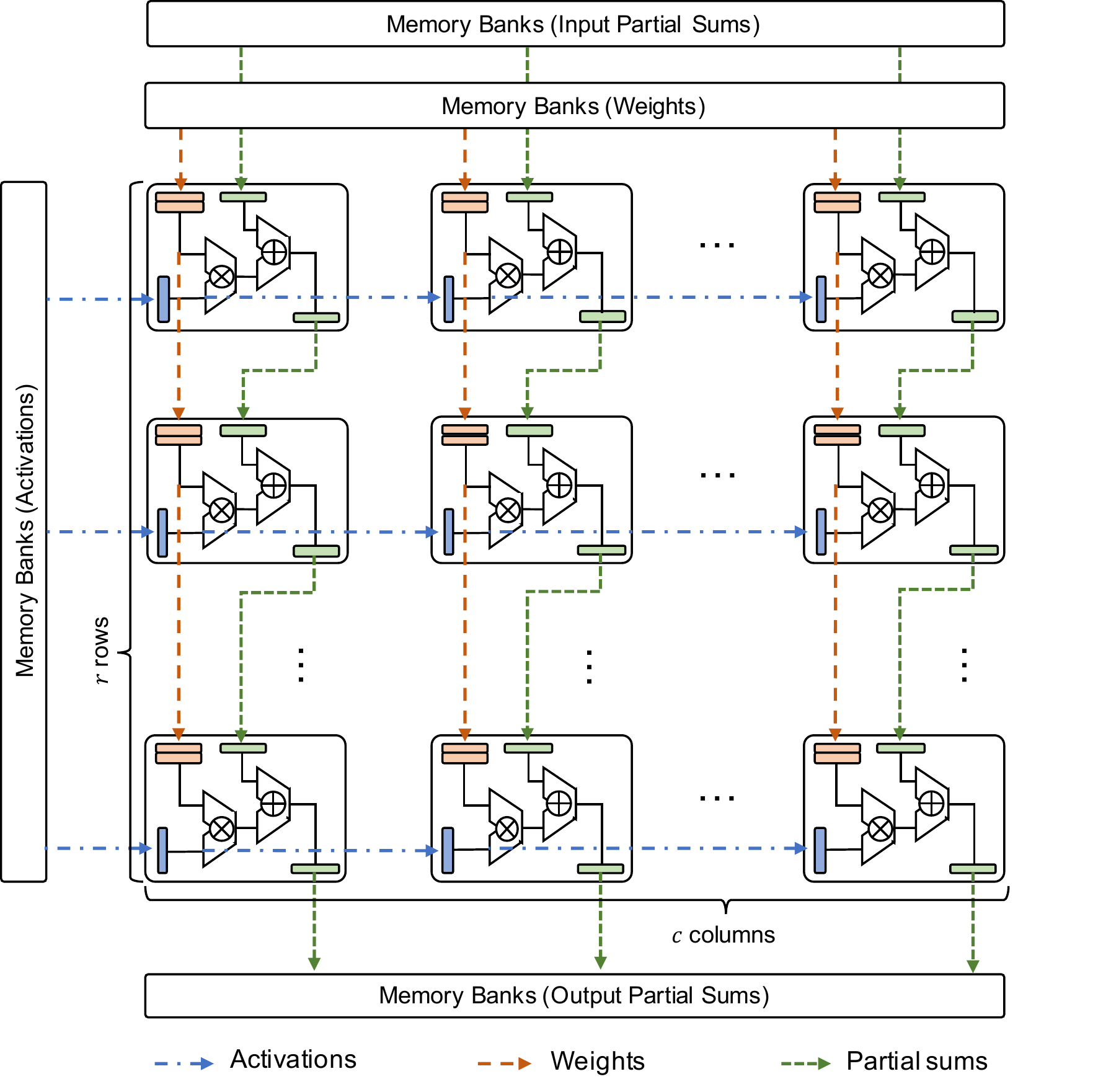}
    	\caption{A weight-stationary systolic array with $r$ rows and $c$ columns. Activations are assumed to traverse along the rows from left to right, weights and partial sums traverse along the columns from top to bottom.}
	    \label{fig:systolic-diagram}
\end{figure}

Among various versions of systolic arrays \cite{kung82}, we focus on the widely adopted weight-stationary design \cite{jouppi2017, kung2019, kung-asplos}. Figure \ref{fig:systolic-diagram} depicts a weight-stationary systolic array, in which the processing elements (PEs) are placed in a grid of $r$ rows and $c$ columns. The PEs are connected to their neighbors along rows and columns through uni-directional point-to-point links. To perform a GEMM operation, a systolic array first fetches the weight matrix row by row and stores them in dedicated registers in PEs. Then, in every cycle, the PEs in the left-most column fetch activations from the memory banks and pass them to the next column. Likewise, the PEs at the top row fetch the input partial sums from the memory banks, perform a multiply-and-accumulate (MAC) operation, and pass the resulting partial sum to the row below. The operation continues with activations flowing from left to right, partial sums flowing from top to bottom, and weights staying stationary. The PEs at the bottom row produce the final results and write them to the memory banks.

In systolic arrays, only the PEs at the edges access the memory banks, whereas those in the middle fetch their operands from their neighbors. Therefore, the number of memory accesses increases linearly with the array dimensions, while the number of MAC operations grows quadratically. As such, large systolic arrays incur relatively lower memory overhead per MAC operation, which improves the overall power efficiency. However, as we increase $r$ and $c$, the array dimensions exceed matrix dimensions in DNN workloads, resulting in idle rows and columns. If the weight matrix dimensions are smaller than the array dimensions, the excess rows and columns become idle, resulting in underutilization. Moreover, in systolic array designs with double buffering \cite{ross18}, the entire array stalls between operations if the matrix multiplication takes fewer cycles than the weight buffering time. Because the execution time of a GEMM operation is equal to the first dimension of the activation matrix $d_1$ (ignoring pipeline latencies), and the weight buffering time is proportional to the number of rows in the array $r$ (assuming weights are fetched row by row), choosing $r > d_1$ also results in underutilization. In short, systolic arrays with large numbers of rows and columns are more likely to suffer from underutilization than small ones due to dimension mismatches.

\begin{figure}[t]
    \centering
    \includegraphics[width=0.7\linewidth]{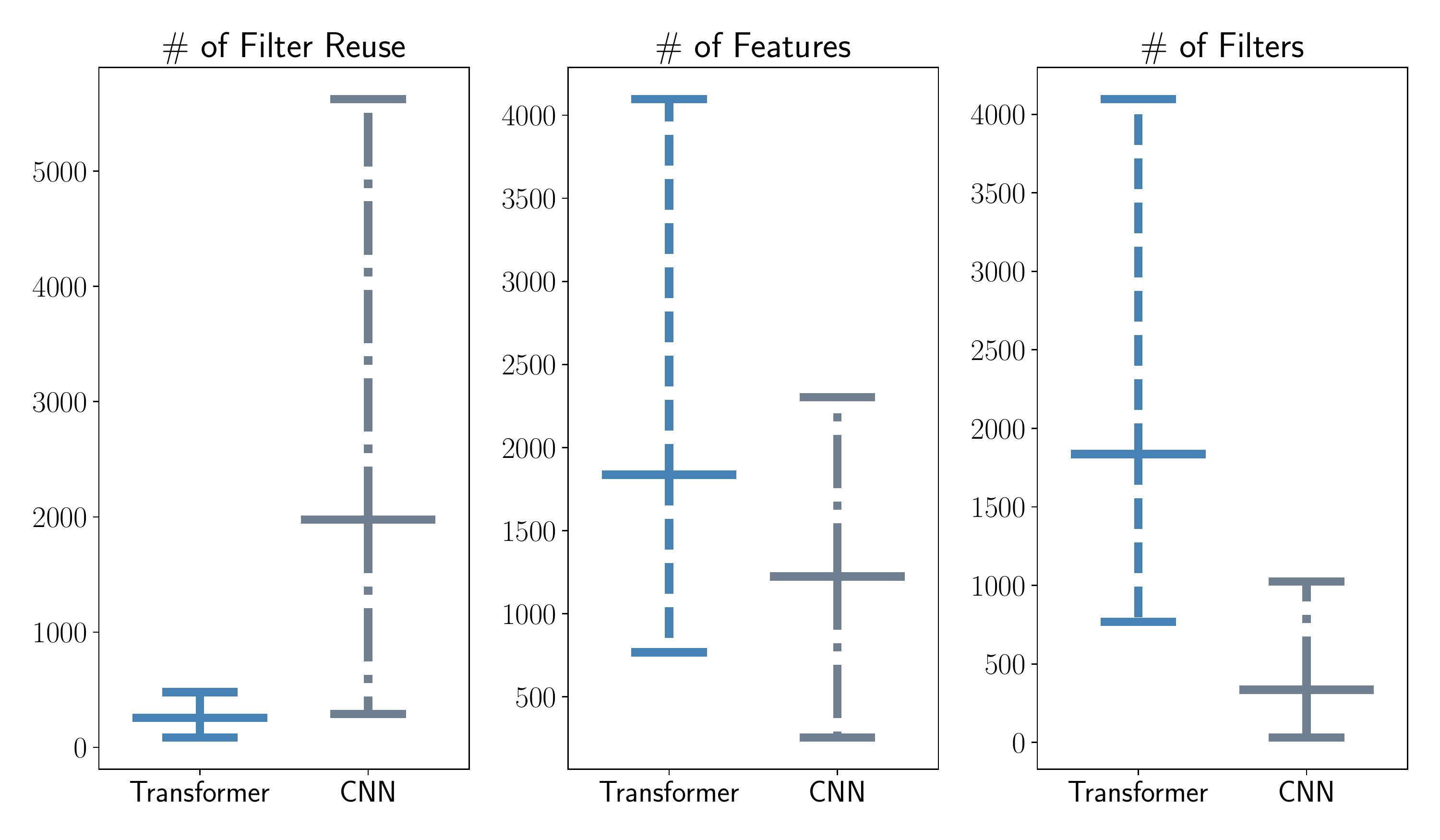}
	\caption{Variation in the number of filter reuse, number of features, and number of filters for the layers of BERT and CNN models. Horizontal lines represent 10th percentile, average, and 90th percentiles, weighted by number of ops in layers.}
    \label{fig:layer-sizes}
\end{figure}

To understand the distinguishing characteristics of different DNN types, we analyze the data dimensions of various state-of-the-art DNN models. Figure \ref{fig:layer-sizes} shows the number of filter reuse (first dimension of $X$), number of features (second dimension of $X$, which is also equal to the first dimension of $W$), and number of filters (second dimension of $W$) of layers from various CNN and Transformer models. Although the dimensions of the DNN layers vary both across and within models, we make the following observations. First, CNN models have significantly ($15 \times$ on average) more filter reuse than Transformer models. This large difference is mainly due to the convolutional reuse, i.e., filters in convolutional layers stride across input images, and each stride corresponds to the amount of filter reuse. Second, while the number of filter reuse in Transformer models is limited to their sequence length (typically a few hundreds), they have significantly more filters (about $6 \times$) on average than convolutional layers. Third, both CNN and Transformer models have large numbers of features, which favors large numbers of rows and columns in systolic arrays.

Although our analysis on data dimensions gives us insights into the contrasting requirements of different DNN types, finding the optimal array size requires a more complex hardware modeling including both utilization and power efficiency. To that end, we devised an optimization metric called effective throughput/Watt, that takes these two variables into account. Due to the utilization component of the effective throughput/Watt metric, the optimal array size is sensitive to the selection of target workloads. Therefore, we conduct a design space exploration for three cases: we first study the CNN and Transformer models separately, and then a mixture of both types. To compare our proposed design choice, we pick four baseline array sizes that represent the majority of designs in industry and academia:  $512 \times 512$, which represents the monolithic systolic array; $256 \times 256$, which represents Google's TPU v1 \cite{jouppi2017}, $128 \times 128$, which represents TPU v2 \cite{tpu2017} and AI-MT \cite{baek2020}, and $8 \times 8$, which represents the Maestro \cite{kung2019}. For this design space exploration, we use our systolic hardware model to obtain the power efficiency and utilization values, and then calculate the effective throughput as peak throughput multiplied by utilization. The correctness of the systolic array model used in this section is validated against the functional simulations of our RTL design. Our design space exploration is isopower because DNN inference accelerators are typically bound by power consumption because of their dense arithmetic formats \cite{sze17}.



\begin{figure}[t]
    \centering
    
    \begin{subfigure}[t]{0.45\textwidth}
    \includegraphics[width=\textwidth]{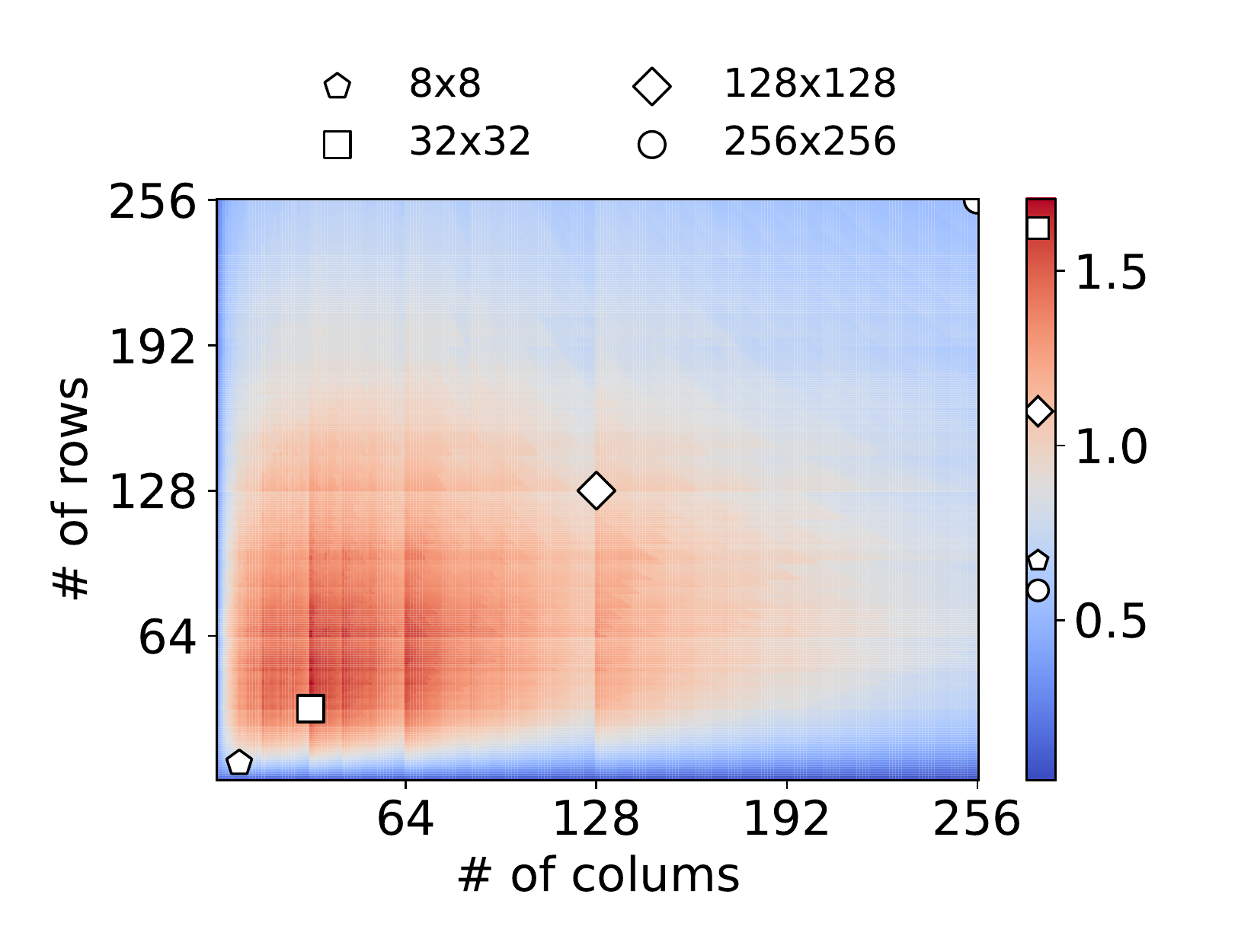}
    \caption{CNN models only.}
    \label{fig:design-space-cnn}
    \end{subfigure}
    \begin{subfigure}[t]{0.45\textwidth}
    \includegraphics[width=\textwidth]{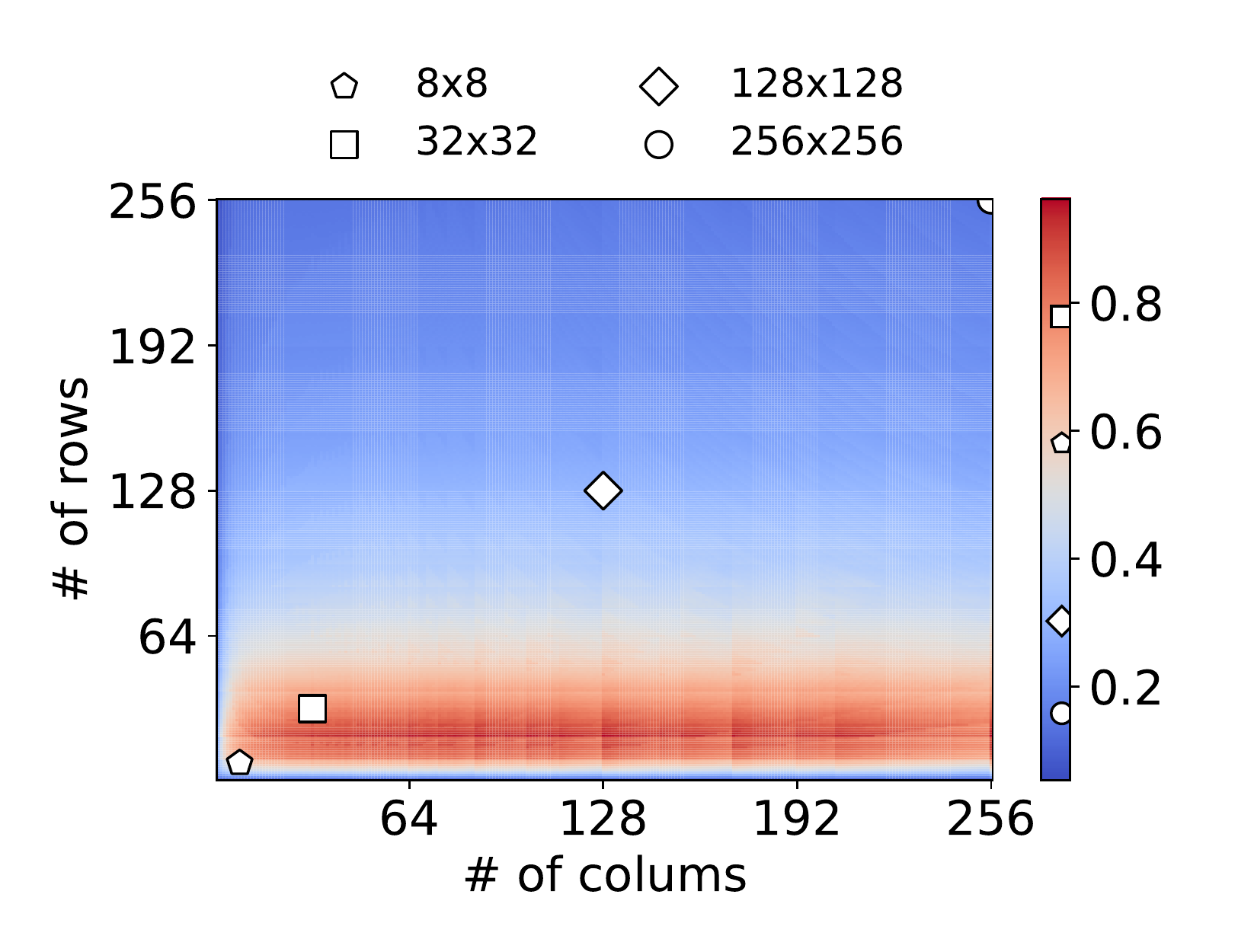}
    \caption{Transformer models only.}
    \label{fig:design-space-transformer}
    \end{subfigure}
    \begin{subfigure}[t]{0.45\textwidth}
    \includegraphics[width=\textwidth]{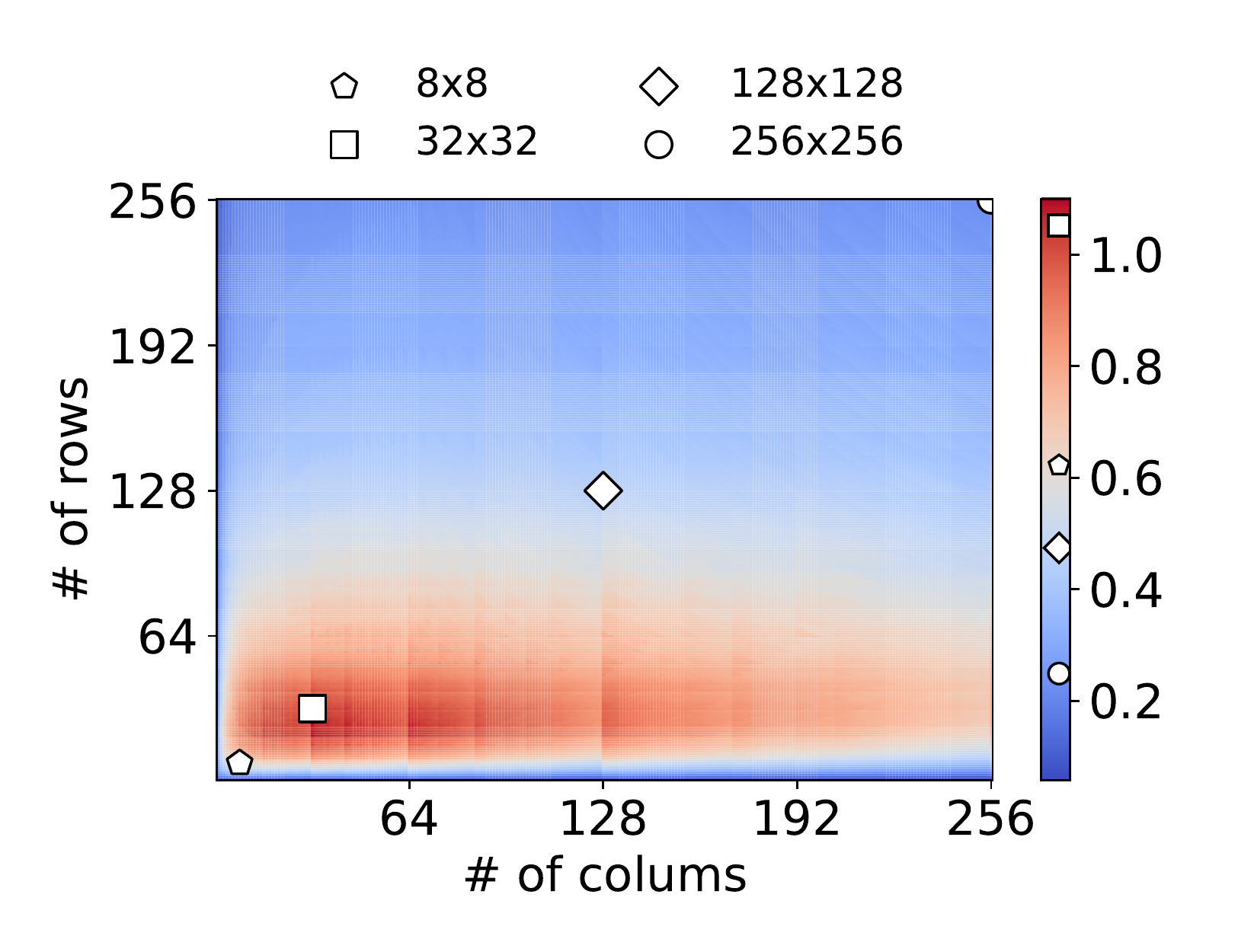}
    \caption{Mix of CNN and Transformer models.}
    \label{fig:design-space-mix}
    \end{subfigure}

	\caption{Design space exploration (isopower) for DNN inference with CNNs only, and Transformers only, and both with equal percentage. The selected CNN models are Inception-v3, ResNet50, ResNet101, ResNet152, DenseNet121, DenseNet169, and DenseNet201 with input image sizes of $224 \times 224$, $256 \times 256$, and $299 \times 299$. The selected Transformer models are BERT-mini, small, medium, base, and large with sequence lengths of 10, 20, 40, 60, 80, 100, 200, 300, 400, 500 as taken from \cite{turbotransformer}. The colormaps represent the effective throughput (TeraOps/s) per Watt, whereas x- and y-axes are the number of columns and rows of a systolic array, respectively. Ripples and discrete lines in the images occur due to the discretization during data tiling.}
	\label{fig:design-space}
\end{figure}

Figure \ref{fig:design-space-cnn} shows the design space exploration for CNN models. Because the number of filters in CNN models is typically limited compared to the number of filter reuse and the number of features as shown in Figure \ref{fig:layer-sizes}, we observe that optimal design points have a large number of rows and a small number of columns. In fact, the design point with the highest effective throughput for CNN models is $66 \times 32$, which is about $1.3 \times$ better than any other baselines. In contrast, Figure \ref{fig:design-space-transformer} shows the design space for Transformer models. Because the number of filter reuse in Transformer models is typically limited compared to the number of filters and number of features as shown in Figure \ref{fig:layer-sizes}, we observe that the optimal design points have a large number of columns and a small number of rows. More specifically, the design point with the highest effective throughput for Transformer models is $20 \times 128$, which is also about $1.7 \times$ better than any other baselines. These two scenarios show that two widely used DNN types, namely CNNs and Transformers, benefit from non-square array shapes, which contradicts the common design patterns in industry or academia. 

Figure \ref{fig:design-space-mix} shows the design space for a scenario that targets both CNN and Transformer models. In this case, data dimensions of both model types have an impact on the shape of the design space: the areas with large array dimensions exhibit low effective throughput/Watt due to underutilization. Likewise, areas with very small dimensions also have low effective throughput/Watt due to poor power efficiency. We identify that only a small part of the design exhibits high effective throughput/Watt, where the number of rows and columns are in the order of ten to hundreds. In fact, the highest effective throughput/Watt is obtained with array dimensions of $20\times32$. Without loss of generality, we choose $32\times32$ as the array size in our design to facilitate implementation and connectivity to memory (e.g., alignment to cache block size).

\subsection{Interconnection Network} \label{sec:interconnect}

For the interconnect between systolic arrays and memory banks, we prioritize four design requirements. First, the interconnect should provide sufficient bisection bandwidth to allow continuous data read and write for all systolic pods simultaneously. Second, the interconnect should offer high combinatorial power with multicasting capability to support as many input-output mappings as possible.
Third, the interconnect should have a latency shorter than the execution of tile operations so that the interconnect latency can be hidden by computation. Finally, the interconnect should scale well up to hundreds of systolic pods in terms of power consumption and silicon area.

2D mesh~\cite{Chen17,Chen14,gao19,shao19} and H-trees~\cite{kung2019,Stevens19} are popularly used in many DNN accelerators thanks to their relatively low hardware cost.
However, neither of them can provide enough bisection bandwidth for large numbers of systolic pods. To improve their bisection bandwidth, one can replicate an H-tree interconnect $N$-times (scaled-up H-tree \cite{kung2019}), which results in an unfeasible hardware cost with a complexity of $N^2$.
Although Crossbar interconnect \cite{Zhu20} offers high bisection bandwidth, it also has a quadratically increasing hardware cost with respect to the number of pods. As such, we conclude that H-tree and Crossbar are not suitable for multi-pod DNN accelerators due to their excessive hardware cost.

Multistage interconnect networks emerge as a promising solution for the energy-efficient multi-pod DNN accelerators as they exhibit sufficient bisection bandwidth, relatively low hardware cost with a complexity of $N \log N$ and low latency.
Prior DNN accelerators\cite{Qin20} have proposed to use Benes network\cite{benes64}, which is a non-blocking multistage interconnection that consists of $(2\log N - 1)$ stages. 
Compared to the H-tree and Crossbar, the Benes network has a feasible hardware cost of $N \log N$ while providing a sufficient bisection bandwidth of $N$. 
However, the Benes network suffers from a considerable latency that is proportional to its large number of stages, $2\log N-1$. 
While the Benes network can route all possible input-output permutations without contention, it offers only a limited multi-casting capability.
Due to this limitation, we consider the augmented version of the Benes network with a copy network \cite{liew2010} instead of its standard implementation, which enables the full multi-casting capability at the expense of longer latency.
Consequently, none of these interconnects mentioned above satisfy the design requirements of an accelerator with a large number of pods.

\begin{figure}[t]
    \centering

    \includegraphics[width=0.4\linewidth]{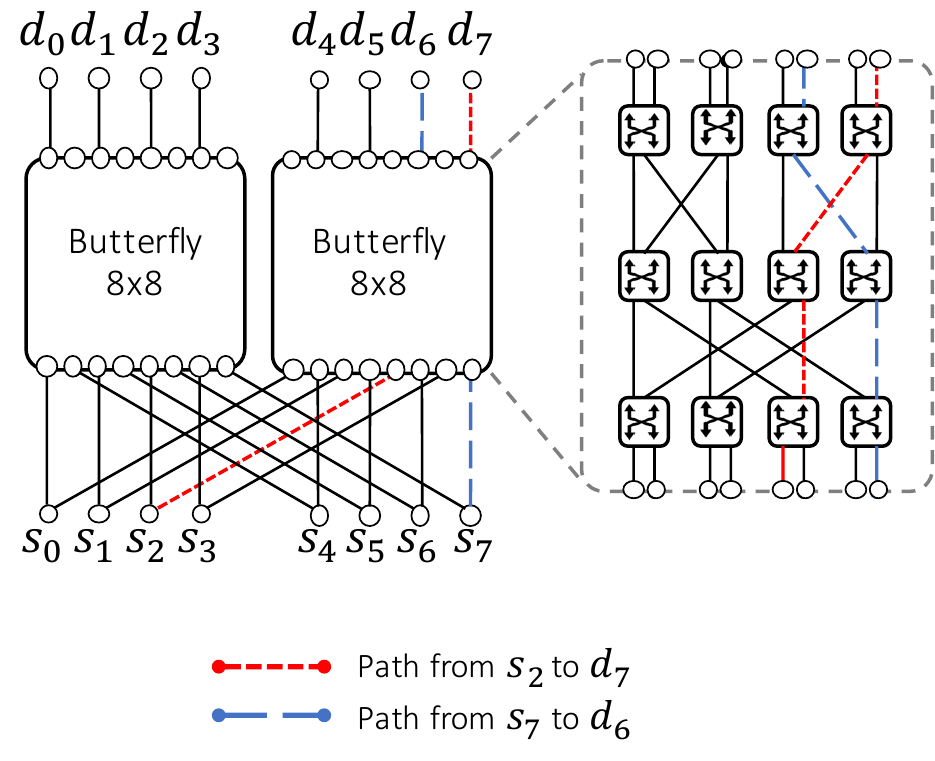}
    \caption{An $8 \times 8$ Butterfly network with an expansion of 2. Routings from $s_2$ to $d_7$ and from $s_7$ to $d_6$ are shown with blue and red lines, respectively.  }
    \label{fig:butterfly}

\end{figure}

The Butterfly network, which is also a multistage inter-connect, offers high bisection bandwidth with low latency and scalable hardware cost. A standard Butterfly network provides only limited multicasting and combinatorial power; however, this limitation can be alleviated by employing multiple of them in parallel, which is referred to as the expansion network \cite{liew2010}.
Figure \ref{fig:butterfly} depicts an example Butterfly network with eight source and destination ports, and with an expansion factor of two. Thanks to the redundant switches and links between source and destinations facilitated by the expansion, we achieve higher combinatorial power than the standard Butterfly network; i.e., more permutations are feasible without network contention. For instance, the paths from $s_3$ to $d_2$ and $s_6$ to $d_3$ can be routed simultaneously as shown in Figure \ref{fig:butterfly}, whereas this permutation would not be possible in a standard Butterfly implementation. Moreover, because the network is expanded vertically rather than horizontally, its latency remains low, allowing to overlap data movement with computation for a larger number of pods. In the rest of this paper, we will refer to the expanded Butterfly network as Butterfly-k, where $k$ is the expansion factor.

\begin{table}[b]
\centering
\begin{tabular}{r|c|c|c}

Type &
\multrowC{Busy Pods\\\unit{\%}} &
\multrowC{Cycles per\\Tile Op.} &
\multrowC{mW/byte}
\\\hline

\multrowR{ Butterfly-1 } &
66.81 &
19.72 &
0.23\\\hline

\multrowR{ Butterfly-2 } &
72.41 &
20.17 &
0.52\\\hline

\multrowR{ Butterfly-4 } &
72.26 &
20.27 &
1.15\\\hline

\multrowR{ Butterfly-8 } &
72.43 &
20.48 &
2.53\\\hline

\multrowR{ Crossbar } &
72.38 &
19.73 &
7.36\\\hline

\multrowR{ Benes } &
72.38 &
30.00 &
0.92

\end{tabular}
\caption{Interconnect performance metrics generated by our cycle-accurate simulator averaged across all the workloads.}
\label{tab:ict_analysis}
\end{table}

To evaluate various interconnect types from the perspective of the design requirements, we identified three performance metrics, which are listed in Table~\ref{tab:ict_analysis}.
First, the percentage of busy pods is defined as the average ratio of busy pods to the total number of pods.
We expect that the interconnects with higher combinatorial power achieve a higher percentage of busy pods due to reduced contention.
Second, the number of cycles per tile operation describes how long it takes for a systolic pod to complete a tile operation.
As we overlap the interconnect accesses with the systolic pod computation, the interconnect latency is typically hidden by the execution time of tile operations.
However, if the interconnect latency is too long, it may become exposed, thereby increasing the number of cycles per tile operation metric.
Finally, Watt/byte of the interconnect characterizes the power consumption.
We seek a smaller Watt/byte value for reducing the overall power consumption.

Table~\ref{tab:ict_analysis} presents the previously mentioned performance metrics for various types of interconnects.
First, we observe that the standard Butterfly network has a reduced percentage of busy pods by 6\% due to its insufficient combinatorial power.
However, expanding it with a factor of two is sufficient to increase its percentage of busy pods to that of interconnects with full combinatorial power, such as Crossbar and Benes.
Second, the interconnect types with low latencies (i.e., Butterfly and Crossbar) results in the minimum number of cycles per tile operation because their latencies are hidden by the computation.
However, Benes network, which has a substantially longer latency incurs a significant overhead (around 50\% more) in the number of cycles per tile operation.
Finally, Crossbar, whose power consumption increases quadratically with the number of arrays, requires $8\times$ and $32\times$ more watts per byte than Benes or standard Butterfly networks for 256 pods.
To conclude, the standard Butterfly, Benes, and Crossbar interconnects are not suitable for multi-pod accelerators due to their insufficient combinatorial power, long latency, and high watts per bytes metrics, whereas Butterfly network with an expansion of two is the optimal design choice thanks to its sufficiently high combinatorial power, relatively short latency, and low watts per bytes.


\subsection{Tiling \& Scheduling} \label{sec:tiling}
Tiling and scheduling play an essential role in maintaining high utilization across large numbers of pods. In this subsection, we first propose a fixed-length tiling strategy with an emphasis on optimal tiling dimensions that maximize utilization across a large number of pods. Then, we explain our scheduler implementation, which maps tile operations onto systolic pods while handling the interconnect constraints, tile dependencies, and bank conflicts.

Performing a GEMM operation on systolic arrays often requires partitioning data into tiles due to dimension mismatches. The resulting tile operations can be performed on a single array sequentially, or they can be distributed among multiple arrays and performed in parallel. In weight stationary systolic arrays, the weight matrix $W$ is spatially laid out onto the systolic arrays; thus, $W$ must be partitioned into tiles of $r \times c$ to match the array dimensions, where $r$ and $c$ denote the number of array rows and columns, respectively. Because the first dimension of $W$ must match the second dimension of the activation matrix $X$ in a matrix multiplication, $X$'s second dimension is also required to be partitioned with a size of $r$. Moreover, we can further partition $X$'s first dimension to obtain more tile operations. Because the execution time of a matrix multiplication is approximately equal to $X$'s first dimension, the resulting tile operations have shorter execution times.

We observe that the data tiling strategies proposed by prior work either do not partition $X$'s first dimension \cite{baek2020}, or choose a partition size that is much larger than array dimensions \cite{Choi20}. We argue that not partitioning $X$'s first dimension or choosing a large partition size does not exploit the available data-level parallelism in matrix multiplication operations to the fullest extent, limiting the number of pods that can run in parallel. Partitioning the $X$ matrix into smaller tiles produces more tile operations that can run in parallel. However, decreasing the partition size below a certain threshold value results in underutilization within pods. This threshold is the number of rows in an array ($r$) because the execution time for tile operations becomes shorter than $r$ cycles, which exposes the weight buffering time. Therefore, we propose to partition the activation matrix $X$ into tiles of $r \times r$, which produces as many parallel tile operations as possible without undermining utilization within pods.

\section{Scale-out Systolic Arrays}
\label{sec:scale-out-arch}

\begin{figure}[t]
    \centering
    \includegraphics[width=1.0\linewidth]{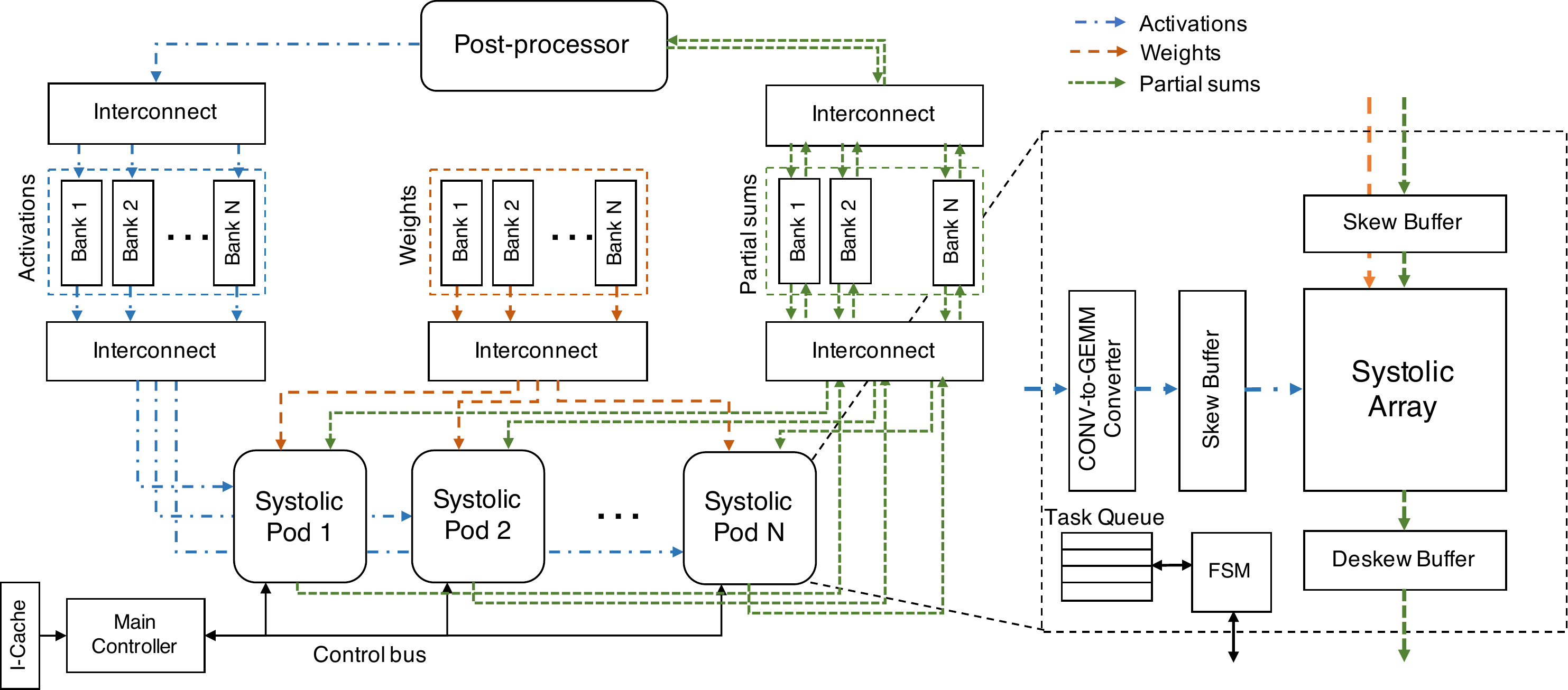}
	\caption{
	Overview of the proposed architecture, with the internals of the Systolic Pod shown on the right-hand side.
	}
    \label{fig:overall-diagram}
\end{figure}

Our baseline SOSA accelerator employs systolic arrays with dimensions ($32 \times 32$) that maximize effective throughput/Watt. In today's server form factors, which allow up to several hundreds of Watts, we can allocate hundreds of optimally sized systolic arrays in parallel to accelerate DNN inference workloads. Unfortunately, employing such large numbers of parallel systolic arrays poses a unique set of challenges to maintain high utilization across arrays. In this section, we also show how the proposed architecture overcomes these challenges.

Figure \ref{fig:overall-diagram} shows the overall diagram of a SOSA accelerator. Each systolic pod encapsulates a systolic array with the peripherals required to perform GEMM operations. The main controller fetches instructions from a dedicated cache, issues them to the corresponding pods, and synchronizes the pods to perform their operations in lockstep. Activation, weight, and partial sum tiles are stored in dedicated on-chip memory banks to reduce the interconnect width between memory banks and systolic pods. A SIMD post-processor performs element-wise operations on the partial sums and writes its results back to the activation or partial sum banks.

\subsection{Systolic Pod Microarchitecture}

Our systolic pod design brings CONV-to-GEMM converter and skew/deskew buffers near systolic arrays to minimize interconnect traffic. As shown on the right-hand side of Figure \ref{fig:overall-diagram}, a systolic pod consists of a systolic array, a CONV-to-GEMM converter, skew/deskew buffers, and a local controller (FSM) with a task queue that stores instructions. The CONV-to-GEMM converter \cite{Liu20b} is a hardware block that converts activation data from a four-dimensional convolutional to two-dimensional matrix format to prevent redundant on-chip memory accesses. Likewise, skew and deskew buffers apply a skew to activation and input partial sums, and remove the skew from output partial sums to improve the memory efficiency. In this design, we encapsulate the systolic arrays with the CONV-to-GEMM converters and skew/deskew buffers to improve memory efficiency and reduce network traffic.

Unlike a standard implementation, we designed our systolic arrays with activation multicast and partial sum fan-in methods \cite{kung82, Liu20a} to reduce pipeline latencies between operations. There are two critical design parameters for these methods: the number of activation multicast (denoted as $U$) and the number of partial sum fan-in (denoted as $V$). In each cycle, the activation values are multicasted to $U$ consecutive processing elements along the rows, while partial sums propagate with an offset of $V$ processing elements along the columns. On the one hand, setting the parameters $U$ and $V$ as one (corresponds to the standard systolic arrays) leads to the best timing thanks to the short paths between registers. However, it incurs large pipeline latencies that result in idle processing elements between tasks. On the other hand, choosing large $U$ and $V$ parameters hinders timing characteristics due to longer paths between registers, but it improves utilization thanks to reduced pipeline latencies. For the optimal array size that we found in our design space exploration ($32 \times 32$), we choose the parameters $U$ and $V$ as 16 and 16, respectively.

\subsection{Offline Scheduling Algorithm}

\begin{figure}[t]
    \centering
    \includegraphics[width=1.0\linewidth]{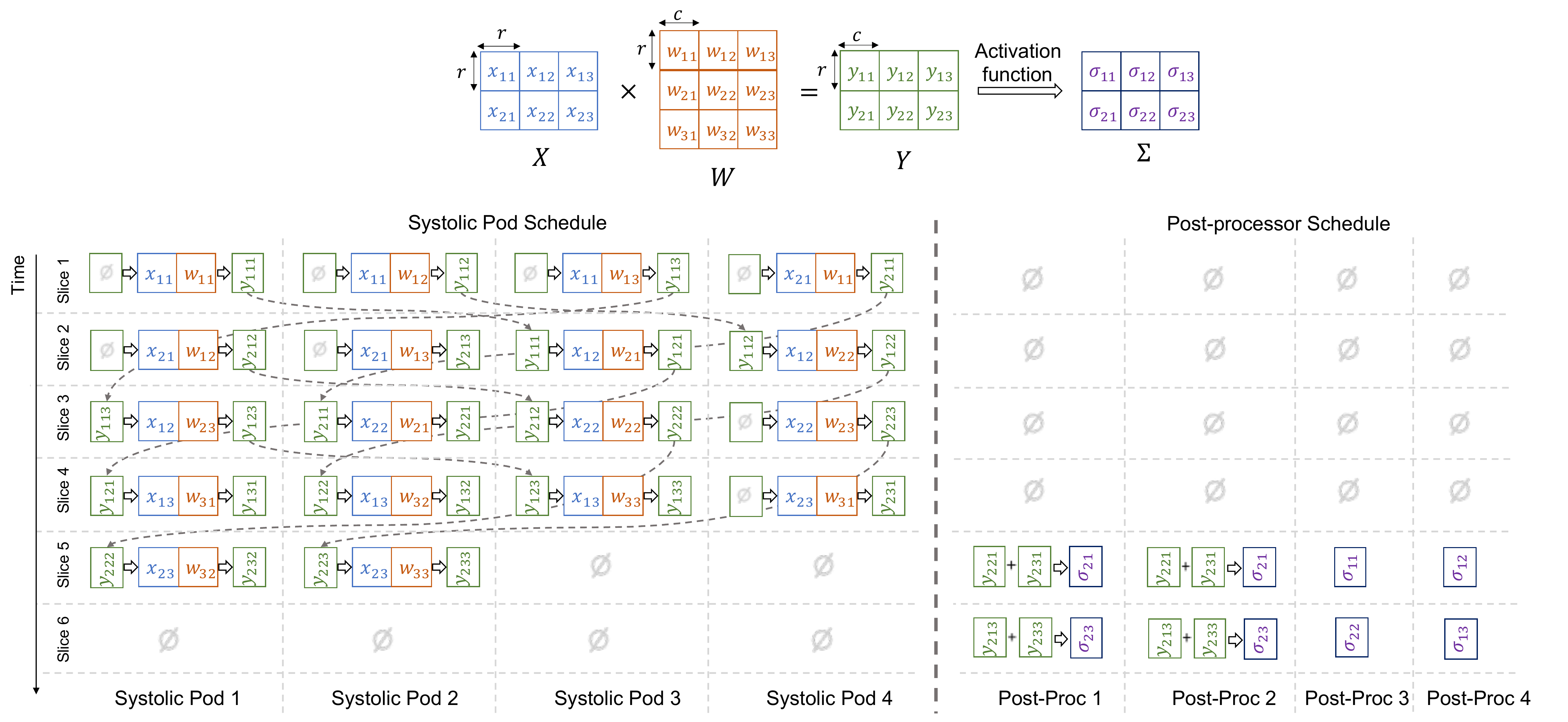}
    	\caption{Tiling and scheduling example of a matrix multiplication of $X \times W \rightarrow Y$, followed by an activation function $Y \rightarrow \Sigma$. The example shows the scheduling for four systolic pods with array sizes of $r \times c$, and four post-processors.}
	    \label{fig:schedule}
\end{figure}

Based on the fixed partition size that we described in Section \ref{sec:tiling}, we propose an offline scheduling algorithm for multi-pod systems. Because our data tiling scheme produces tile operations with identical execution times, we design a scheduler with fixed time slices of $r$ cycles. The scheduler takes a list of tile operations generated by the tiling algorithm and attempts to map and schedule them on the slots of available systolic pods at the earliest available time slice. The scheduler has three constraints while checking a tile operation's availability for a slot. First, there must be no read-after-write data dependence between scheduled tile operations. Second, a memory bank can not be accessed by another pod as we assume single-ported banks. Third, the interconnect must be able to route all pod-bank permutations for a given time slice. Starting from the first tile operation, the scheduler searches for available slots in time slices that satisfy all three conditions.

The scheduler first finds the earliest possible time slice ($l$) by checking the dependencies among tile operations. Then, it finds the idle systolic pods and memory banks in the time slice found in the previous step. Next, it exhaustively searches all combinations of available pods and memory banks and checks whether a routing between pods and banks is possible. If it finds a valid routing for all $X$, $W$, and $P$ interconnects, it schedules the tile operation in the time slice $l$. If it fails to find any valid routing after all combinations are exhausted, it repeats the same steps for the next time slice, $l+1$. The scheduler repeats this process until all tile operations are scheduled. 

Figure \ref{fig:schedule} shows the result of our tiling and scheduling algorithm on a small-scale example. Each column in the figure represents a systolic pod or post-processor and each row represents a time slice. In each time slice, a systolic pod performs a tile operation $x_{ij} \times w_{jk} + y_{imk} = y_{ijk}$, where $x_{ij}$ and $w_{jk}$ are tiles from $X$ and $W$, respectively, $y_{imk}$ is an optional input partial sum, and $y_{ijk}$ is the output partial sum. The output partial sums ($y_{ijk}$) are then aggregated to obtain the final output tiles: $y_{ik} = \sum_j y_{ijk}$. Finally, post-processors applies an activation function on final output tiles ($y_{ik}$) to obtain output activations, $\sigma_{ik}$.

Due to the aggregation operations between partial output tiles, there are data dependencies between the tile operations. There are two ways of performing these tile aggregations. The output of a tile multiplication can be mapped onto a later multiplication as the input partial sum (shown with dashed lines in Figure \ref{fig:schedule}), or the aggregation of two tiles can be performed in the idle slots of post-processors. Post-processors work in pairs to perform tile aggregations to match the throughput of systolic pods. We use an exhaustive search to map aggregation operations on systolic arrays and post processors.

\section{Methodology} \label{sec:methodology}

We synthesized the proposed systolic pods using the TSMC 28nm process technology and Synopsis Design Compiler, then measured that the energy consumption per MAC operation is 0.4 pJ at a clock frequency of 1GHz. We encoded the weight and activation values as 8-bit and partial sums as 16-bit integers, the same as prior work \cite{drumond20, jouppi2017, baek2020}. We modeled the on-chip memory banks using Cacti-P \cite{li11}. As we use an N-to-N interconnect, we employ the same number of SRAM banks as the number of systolic pods. We chose the SRAM bank size as 256 KB, which is the smallest bank size that can store the working set of all of the benchmarks. We calculated that the energy per byte for accessing memory banks is 2.7 pJ/Byte using Cacti-P \cite{li11}. For off-chip memory access, we assume HBM memory as in TPUv3 \cite{tpu2017}.

To evaluate the proposed method, we selected a number of benchmarks from the two most widely used application domains of DNNs, namely computer vision and natural language processing. Because the majority of DNN models (all top ten models in the Imagenet competition) for computer vision tasks are convolutional, we choose a number of widely used, state-of-the-art CNN models, namely Inception-v3 \cite{szegedy16}, ResNet50, ResNet101, ResNet152 \cite{he16}, DenseNet121, DenseNet169, and DenseNet201 \cite{Huang17}. We use the pre-trained models provided by Keras \cite{keras} for CNNs with an input image size of $299 \times 299 \times 3$. Transformer models are ubiquitous in the NLP domain (eight out of the top ten in the WMT-14 English-German dataset use a form of Transformer models). Therefore, we select three BERT models \cite{devlin19}, namely BERT-medium, BERT-base, and BERT-large \cite{bert}. For BERT models, we select the median value of the sequence lengths from the benchmark \cite{turbotransformer}, which is equal to 100.

\section{Results}

In this section, we first show that the proposed array size ($32 \times 32$) offers the highest effective throughput among all other design points. Then, we show and discuss the impact of multi-batching and multi-tenancy in the effective throughput and its scalability. After that, we evaluate various interconnect types presented in Section \ref{sec:scale-out-arch} and demonstrate that the Butterfly network is the most optimal choice to connect large numbers of systolic pods. Next, we evaluate the proposed tiling scheme and quantitatively show the improvement achieved by choosing the optimal tiling size. Later, we analyze the SRAM bank size's impact on the off-chip DRAM usage and effective throughput, and identify the optimal SRAM bank size for the proposed design. Finally, we share the details of our RTL implementation and synthesis results to show the validity of our insights and assumptions. For all design points, we consider the thermal design power (TDP) of 400 Watts, which is taken from \cite{nvidia-a100}, and calculate the number of parallel systolic arrays as the largest power-of-two number that results in a peak power consumption smaller than the TDP.

\subsection{Array Granularity}

\begin{table}[b]
\centering
\begin{tabular}{ r | r | r | r | r | r}

\multrowC{Systolic Array\\Size} &
\multrowC{\# of\\Pods} &
\multrowC{Peak\\Power\\\unit{Watts}} &
\multrowC{Peak\\Throughput\\@400W\\\unit{TeraOps/s}} &
\multrowC{Util.\\\unit{\%}} &
\multrowC{Effective\\Throughput\\@400W\\\unit{TeraOps/s}} \\\hline


$512 \times 512$ & 1 & 113.2  & 1853 & 10.3 & 191.3 \\ 
$256 \times 256$ & 8 & 245.0 & 1712 & 14.0 & 183.0 \\ 
$128 \times 128$ & 32 & 283.1 & 1481 & 13.8 & 205.0 \\ 
$64 \times 64$ & 128 & 362.2 & 1158 & 17.4 & 200.9 \\ 
$16 \times 16$ & 512 & 210.6 & 498.0 & 40.0 & 198.9 \\ 


$\textbf{32} \times \textbf{32}$ &
\textbf{256} &
\textbf{260.2} &
\textbf{806.0} &
\textbf{39.4} &
\textbf{317.4} \\ 

\end{tabular}

\caption{Performance of SOSA with various array sizes.}
\label{tab:results}
\end{table}

Systolic array designs in academia and industry covers a wide range of array sizes (e.g., $8\times8$ in Maestro \cite{kung2019},  $128 \times 128$ in TPUv2, v3, v4 \cite{tpu2017} and AI-MT \cite{baek2020}, and $256 \times 256$ in TPUv1 \cite{jouppi2017}. To cover the entire design space, we run our experiments with array sizes from $16\times16$ to $512\times512$ with steps of power-of-two numbers. Moreover, to compare SOSA against the monolithic designs (e.g., TPUv1 \cite{jouppi2017}), we have the Monolithic baseline, in which available processing elements are organized in the form of a single systolic array.

Table \ref{tab:results} summarizes the performance results of SOSA with varying array granularities. Because large systolic arrays have higher power efficiency than smaller ones, Monolithic baseline achieves the highest theoretical peak throughput (1.85 PetaOps/s) among all other array granularities. However, due to underutilization in large systolic arrays, Monolithic baseline exhibits only 10.3 \% of its peak throughput, which corresponds to an effective throughput of 191.3 TeraOps/s. In contrast, array granularity of $16 \times 16$ has the lowest peak throughput due to its low power efficiency. As a result, even though it achieves the highest utilization, its effective throughput is only 198.9 TeraOps/s because of its limited peak throughput. In parallel with the purpose of the proposed array granularity, the array size of $32 \times 32$ finds a balance between power efficiency and utilization and maximizes the effective throughput. As a result, the proposed architecture with the array size of $32 \times 32$ achieves the best effective throughput (317.4 TeraOps/s), which is $1.5\times$ higher than all other design points.

\begin{figure}[t]
    \centering
    \includegraphics[width=1.\linewidth]{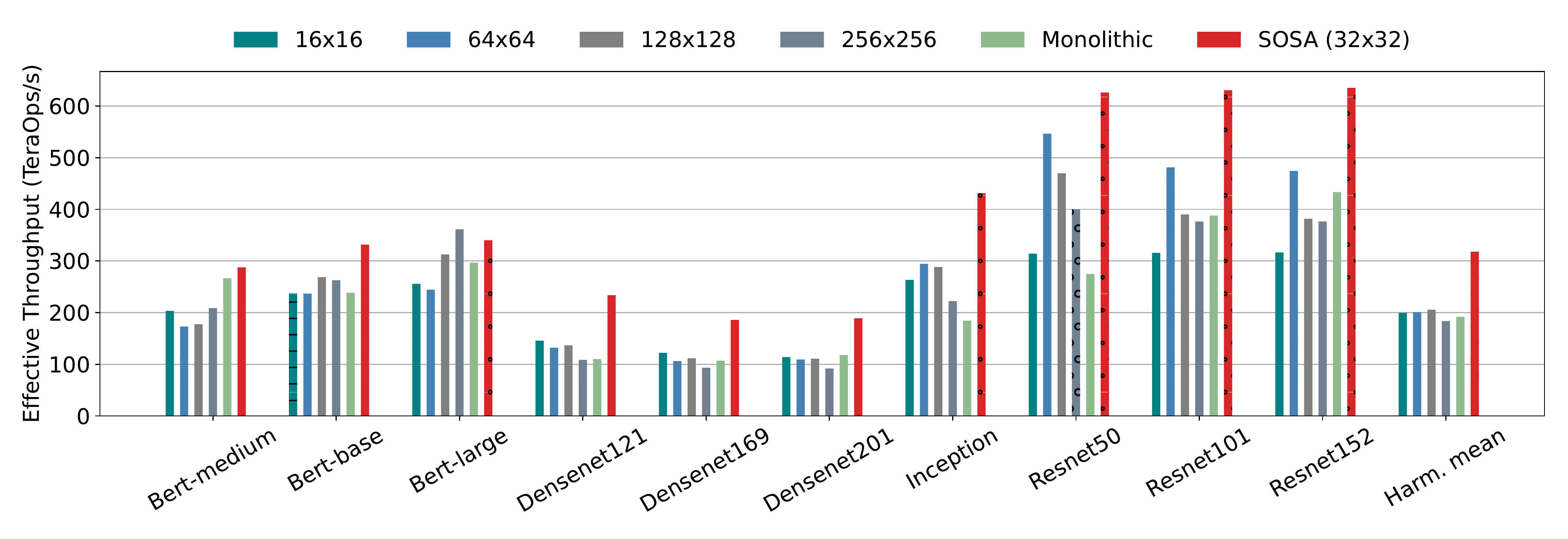}
    	\caption{Effective throughput of SOSA with various array sizes and Monolithic baseline for various DNN benchmarks. All values are normalized to 400 Watts.}
	    \label{fig:results}
\end{figure}

Figure \ref{fig:results} shows the breakdown of effective throughputs for individual DNN models. We observe, in nine out of ten benchmarks, SOSA with the array size of $32 \times 32$ outperforms other designs by up to a factor of $\sim1.6$. The only benchmark that $32 \times 32$ does not exhibit the highest throughput is BERT-large, for which the array size of $256\times256$ outperforms $32 \times 32$ by a factor of $1.06$. We argue that the reason why $256\times256$ outperforms other designs for BERT-large is because its array dimensions are well-aligned with the data dimensions in BERT-large. Nevertheless, these results show that SOSA with the array size of $32 \times 32$ is the optimal design point for a wide range of state-of-the-art DNN workloads, with an average effective throughput higher than prior designs by a factor of $1.55\times$.

\begin{figure}[t]
    \centering
    \includegraphics[width=0.5\linewidth]{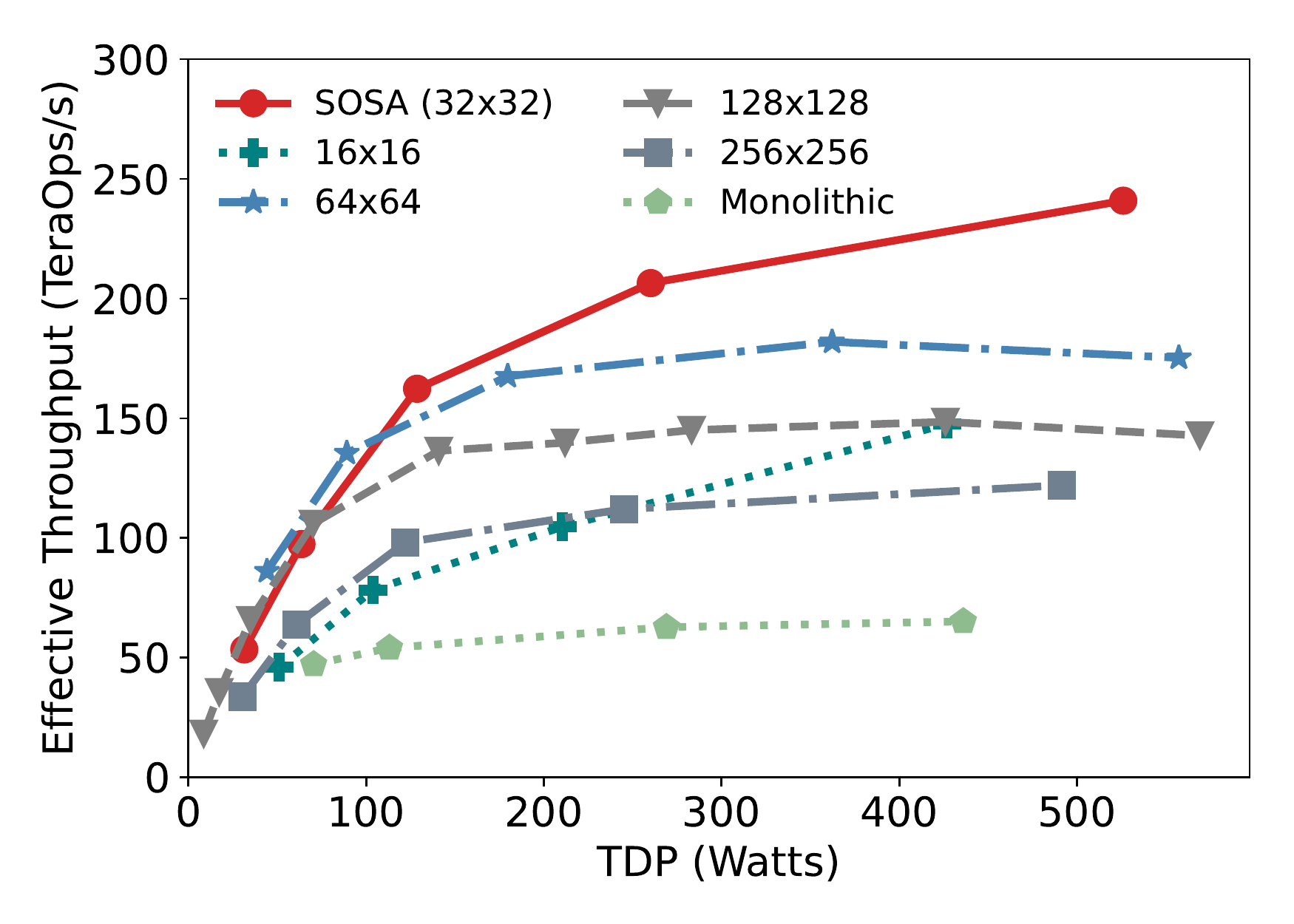}
    	\caption{Effective throughput of SOSA and Monolithic baseline for various TDP values. For Monolithic baseline, we assume a single systolic array and vary its dimensions between $400 \times 400$ and $1024 \times 1024$; whereas for SOSA designs, we use keep the array size constant and vary the number of parallel pods.}
	    \label{fig:perf-scale}
\end{figure}

In the previous evaluation, we assumed the number of systolic pods as 256. However, AI accelerators' power and area budgets may vary depending on the system requirements; thus, we evaluate the proposed architecture's sensitivity to the number of pods. Figure \ref{fig:perf-scale} shows the effective throughput for various numbers of pods. We observe that, for TDP values larger than 90 Watts, SOSA with the array size of $32\times32$ outperforms all other designs by a factor up to $1.5\times$, when the number of arrays is scaled up to 512. We also observe that the increase in the effective throughput starts to saturate as we increase the number of arrays more than 128. This is because of the fact that we target a batch size of one for all workloads to mimic an online setting, which easily falls short of tile operations that run in parallel in large numbers of arrays. This limitation can be easily avoided by increasing the data size, either by increasing the batch size or running multiple workloads in parallel.

\begin{figure}[t]
    \centering
    \includegraphics[width=0.5\linewidth]{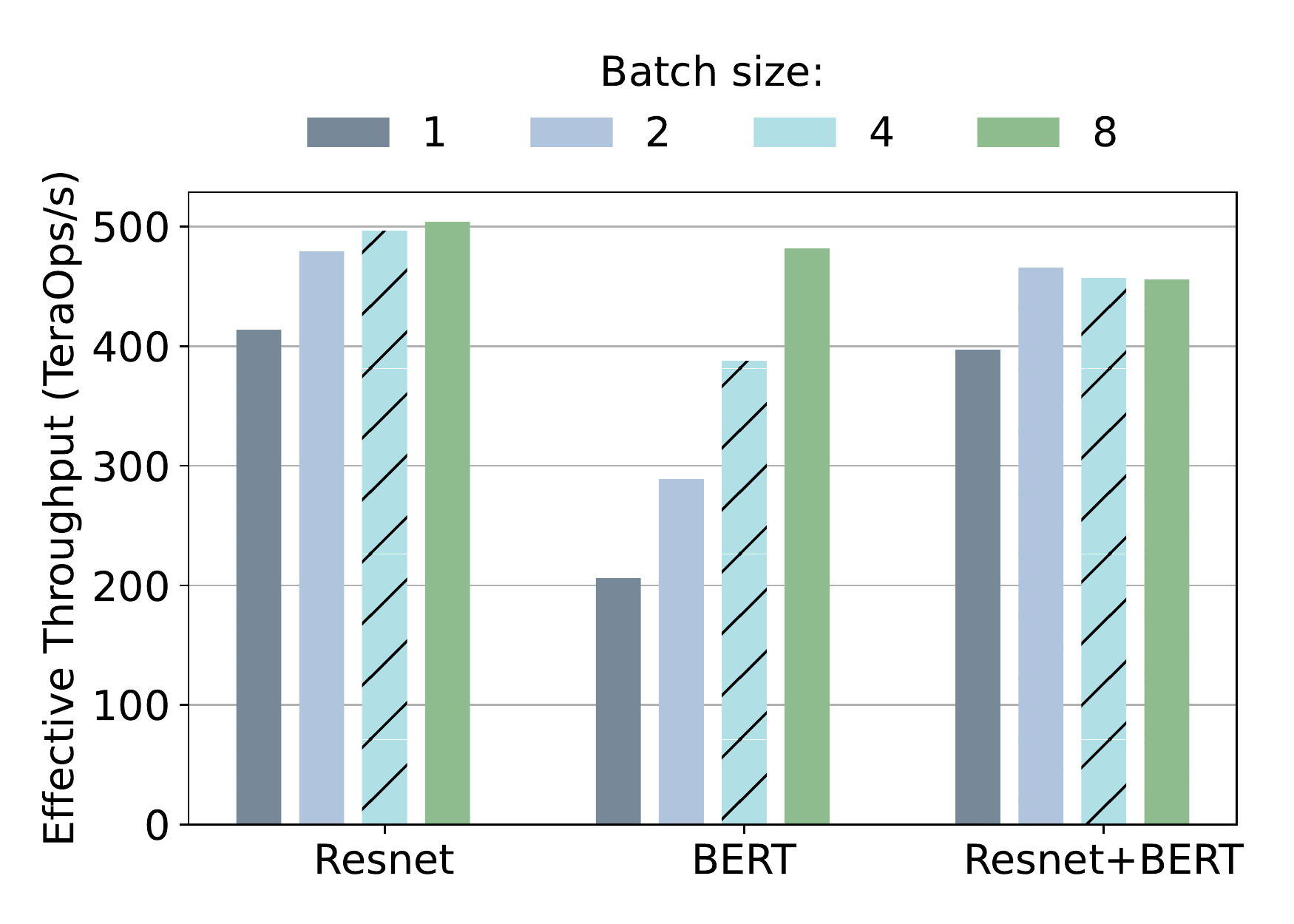}
    	\caption{Effective throughput of SOSA varying batch sizes for Resnet only, BERT only, and both Resnet and BERT in parallel.}
	    \label{fig:multibatch}
\end{figure}

To show the impact of choosing larger batch sizes and running multiple workloads in parallel, we measured the effective throughput of a Resnet-152 and BERT-medium models with varying batch sizes, which is shown in Figure \ref{fig:multibatch}. We observe that, because the proposed design already achieves an effective throughput close to its peak throughput for the Resnet model, increasing the batch size does not lead to a significant improvement in its effective throughput. In contrast, because the proposed design is underutilized while running the BERT model due to insufficient number of tile operations, increasing the batch size results in a significant improvement in effective throughput. Likewise, running multiple workloads in parallel also increases the number of tile operations. As such, running the Resnet and BERT models in parallel with a batch size of one achieves an effective throughput of 397 TeraOps/s, which is $1.44\times$ higher than running them sequentially.

\begin{figure}[t]
    \centering

    \begin{subfigure}[t]{0.45\textwidth}
        \includegraphics[width=\textwidth]{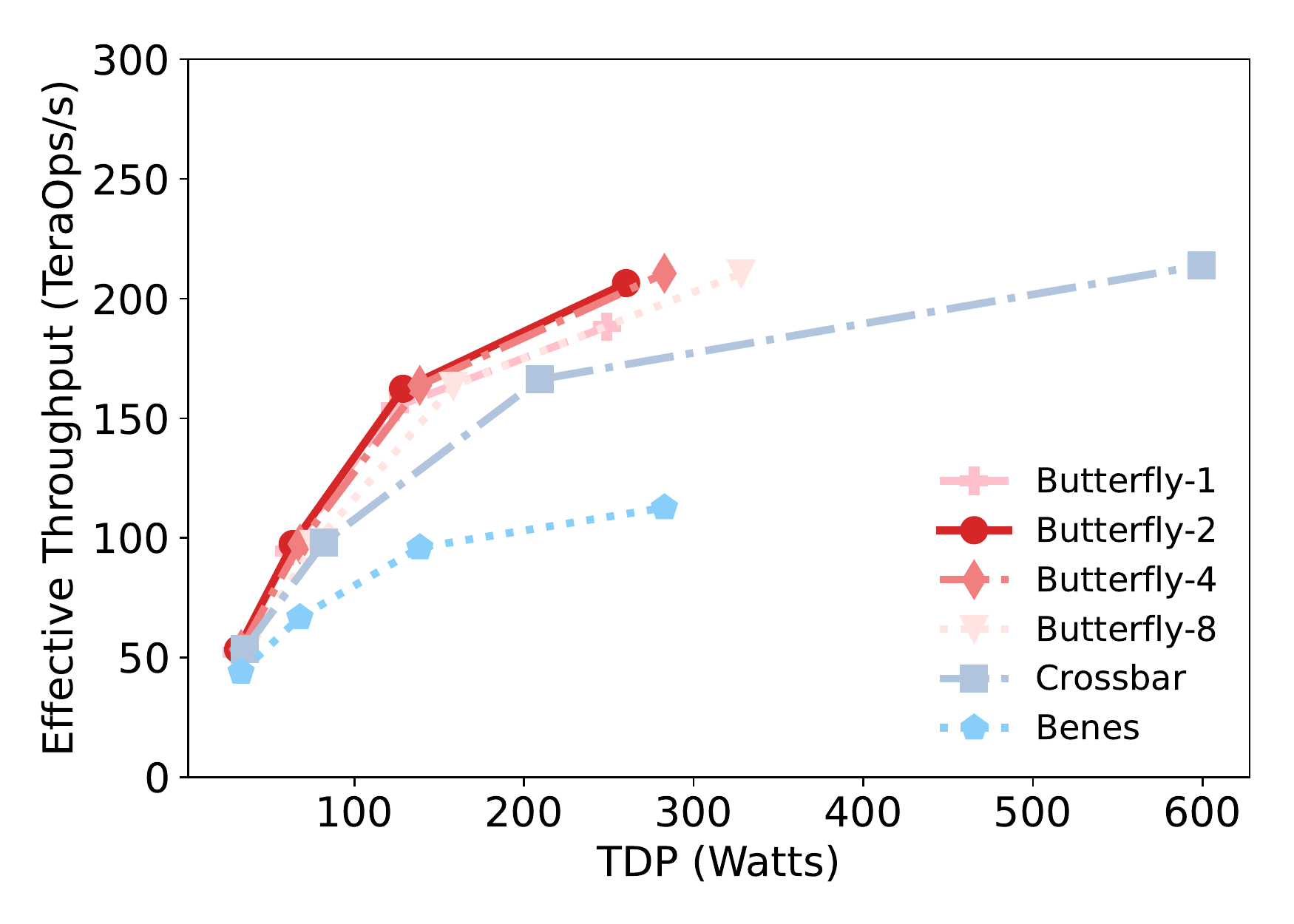}
        \caption{Effective throughput versus TDP for various interconnect types. The number of pods is varied between 32 and 256, and the TDP values are calculated accordingly.}
        \label{fig:interconn-scale}
    \end{subfigure}
    \begin{subfigure}[t]{0.45\textwidth}
        \includegraphics[width=\textwidth]{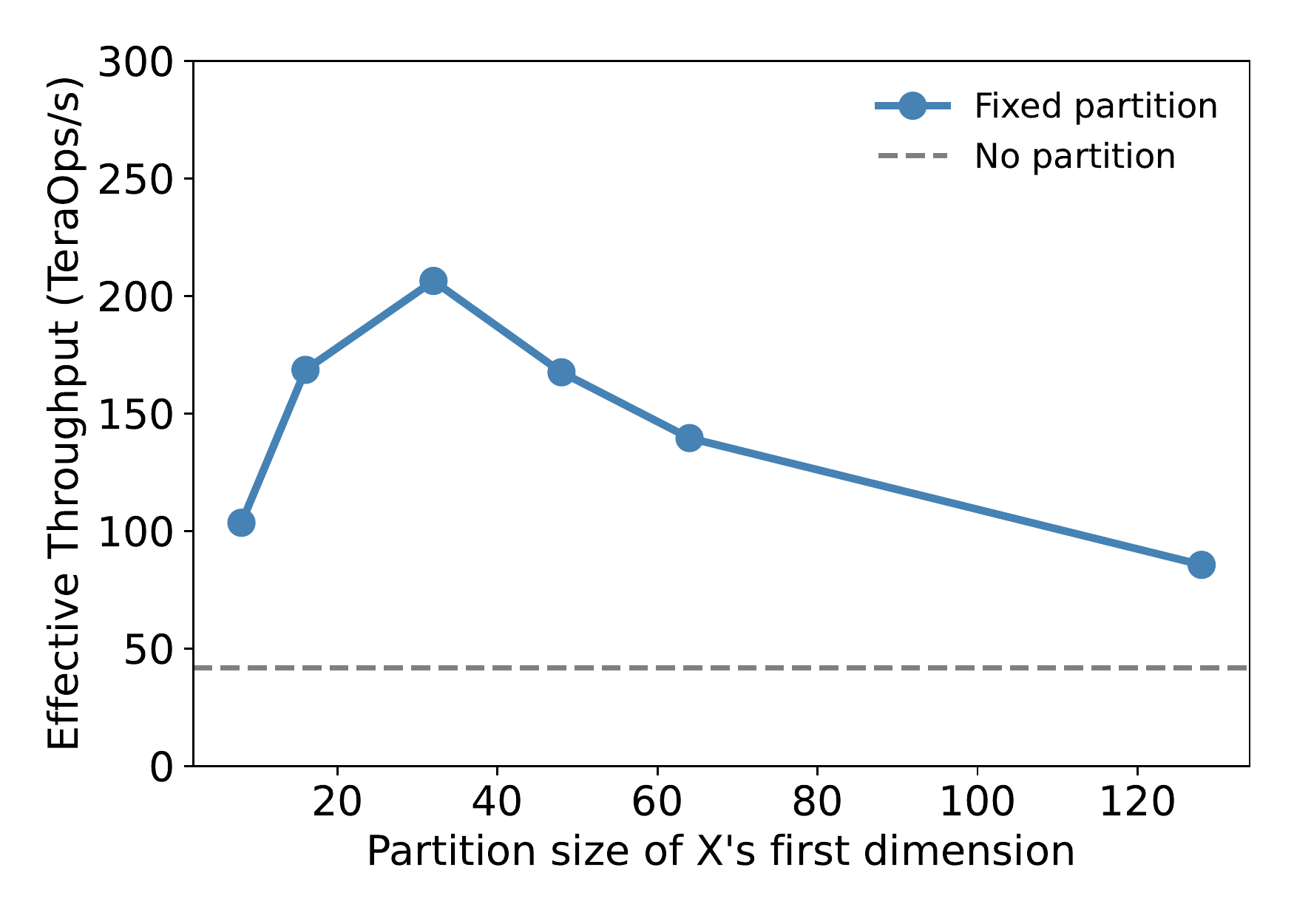}
        	\caption{Normalized effective throughput for varying $k$ values.}
    	    \label{fig:split-size}
    \end{subfigure}
\end{figure}

\subsection{Interconnect}

To demonstrate the interconnect's impact on a multi-pod system, we measured the effective throughput and calculated the TDP for various numbers of pods and for various types of interconnects, which is shown in Figure \ref{fig:interconn-scale}. Crossbar achieves the highest effective throughput (213.8 TeraOps/s) thanks to its high connectivity and low latency but it requires around $2.3 \times$ more peak power than any other interconnect due to its quadratically increasing hardware cost. Although Benes network offers high connectivity, it performs poorly for the increasing number of pods mainly due to the fact that their long latency becomes exposed, reducing its effective throughput. This experiment confirms that Butterfly is the optimal interconnect for multi-pod systems, which achieves an effective throughput only 4\% less than Crossbar but at a much lower TDP.

Figure \ref{fig:interconn-scale} also evaluates the impact of the expansion factor for the Butterfly network. For increasing expansion factors, Butterfly networks are capable of routing more input and output permutations, reducing network contention and improving effective throughput. However, the hardware cost of the interconnect increases with the expansion factor, which reduces its effectiveness. Our experiment shows that the expansion factors larger than two achieve only a marginal increment in effective throughput (less than 2\%) while the hardware cost of the interconnect doubles with every expansion. Thus, we conclude that Butterfly network with an expansion factor of two is the optimal choice of interconnect, with an effective throughput of 206.5 TeraOps/s at a TDP of 260 Watts.

\subsection{Tiling}

To demonstrate the effect of our proposed tiling strategy quantitatively, we measured the effective throughput for CNN and BERT benchmarks using various partition sizes for activation matrix's first dimension (denoted as $k$). Figure \ref{fig:split-size} shows the sensitivity in effective throughput with respect to partition size, $k$. When the activation matrix's first dimension is not partitioned or the partition size is larger than the number of rows in the systolic array, the effective throughput is suboptimal because the number of tile operation that can run in parallel is not sufficient to keep large numbers of arrays active. When $k$ is smaller than the number of rows, the effective throughput also decreases because the interconnect and buffering latencies become exposed. Therefore, we identify the optimal partition size for the activation matrix as $r \times r$, which maximizes the overall effective throughput in a multi-pod system.

    


\begin{figure}[t]
    \centering
    \includegraphics[width=0.5\linewidth]{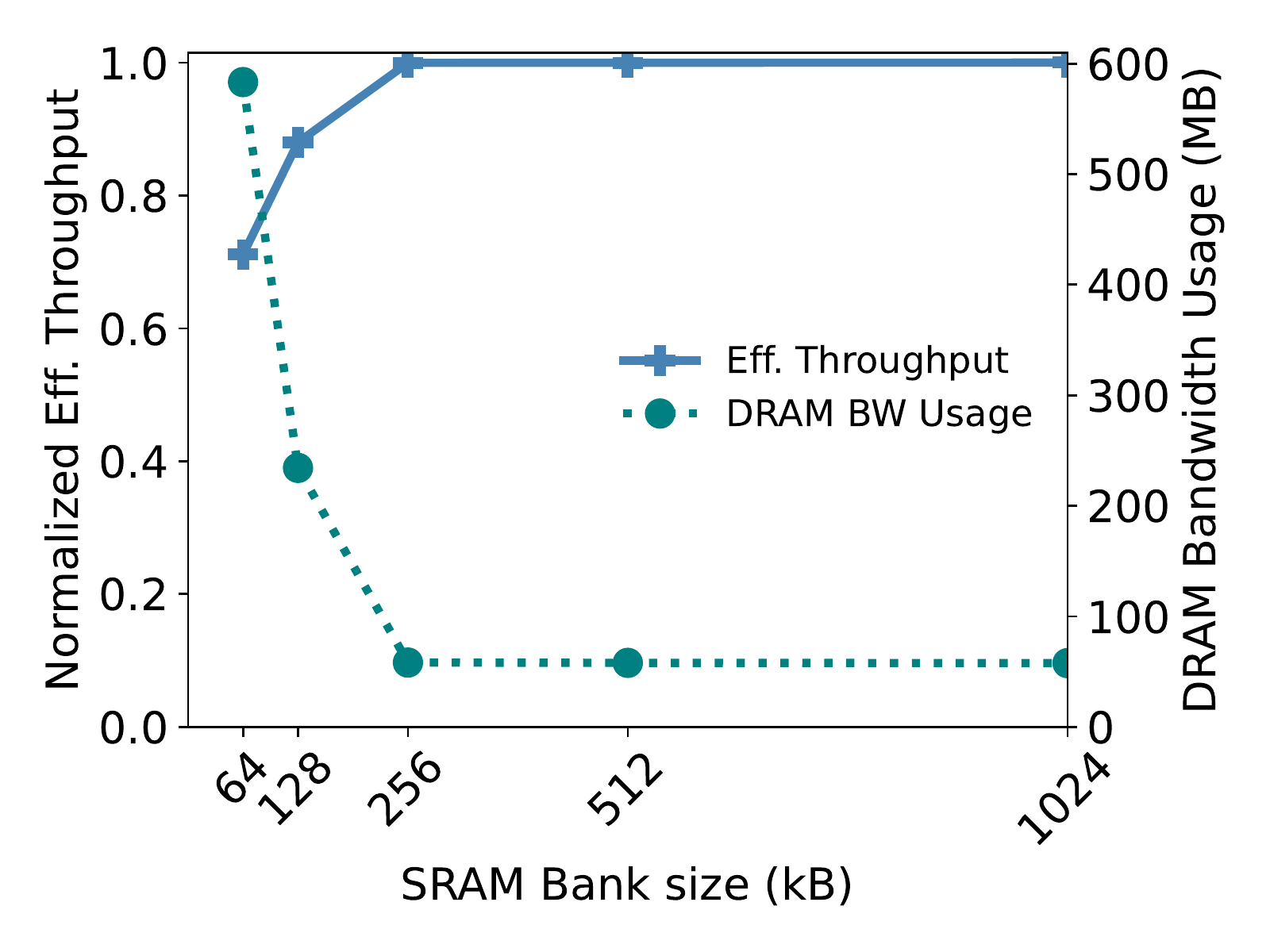}
    	\caption{Effective throughput (normalized to maximum value) and off-chip DRAM usage for varying on-chip SRAM bank sizes.}
	    \label{fig:bank_size}
\end{figure}

\subsection{Memory}
The effective throughput of the proposed architecture is also sensitive to the on-chip SRAM capacity because an SRAM capacity that is smaller than the workload's active working set may lead to latency and energy overhead due to frequent off-chip DRAM accesses, whereas an unnecessarily large SRAM capacity increases energy consumption and silicon area cost. To show the effective throughput's sensitivity to on-chip SRAM capacity, we vary the bank size from 64kB to 1MB, and measured the effective throughput and DRAM bandwidth usage, which are shown in Figure \ref{fig:bank_size}. In this experiment, we used the workload with the largest active working set among all benchmarks, namely Resnet152, with a batch size of eight. We observe that, choosing an SRAM bank size smaller than 256kB results in tile eviction and SRAM misses, which leads to both increased DRAM bandwidth usage, which consequently reduces the effective throughput. As a result, we pick a bank size of 256kB in our proposed design.

\subsection{RTL Synthesis}

To verify our assumptions on hardware parameters, we synthesized the proposed design in Synopsys Design Compiler using TSMC 28nm library. Table \ref{tab:synthesis-results} summarizes our synthesis results. We observe that on-chip SRAM memory constitutes the majority of the power consumption ($\SynthPercentagePowerSRAM\%$) and silicon area ($\SynthPercentageAreaSRAM\%$). The results also demonstrate that the proposed interconnect, namely the Butterfly network with an expansion factor of two, is only $\SynthPercentagePowerInterconnect\%$ of the total power consumption and $\SynthPercentageAreaInterconnect\%$ of the total silicon area. Unsurprisingly, systolic arrays constitute a large portion of a pod's total power consumption ($\SystolicArrayPowerRelativePod\%$) and silicon area ($\SystolicArrayAreaRelativePod\%$), whereas the rest is taken by the control logic and buffers.

\begin{table}[h!]
\centering
{
\begin{tabular}{c|l|r|r}

\multicolumn{2}{l|}{} &
Power \unit{\%} &
Area \unit{\%}
\\\hline


\multicolumn{2}{l|}{SRAM} &
\SynthPercentagePowerSRAM &
\SynthPercentageAreaSRAM
\\\hline

\multicolumn{2}{l|}{Post-processor} &
\SynthPercentagePowerPPFX &
\SynthPercentageAreaPPFX
\\\hline

\multicolumn{2}{l|}{Interconnect} &
\SynthPercentagePowerInterconnect &
\SynthPercentageAreaInterconnect
\\\hline

\multirow{7}{*}{\begin{tabular}[c]{@{}c@{}}\rotatebox[origin=c]{90}{Systolic Pod}\end{tabular}}

&
Systolic Array &
\SynthPercentagePowerPodArray &
\SynthPercentageAreaPodArray
\\\cline{2-4} 

&
Job Queue &
\SynthPercentagePowerPodJobQueue &
\SynthPercentageAreaPodJobQueue
\\\cline{2-4} 

&
Act. Buffer &
\SynthPercentagePowerPodActBuffer &
\SynthPercentageAreaPodActBuffer
\\\cline{2-4} 

&
Conv. Buffer &
\SynthPercentagePowerPodConvBuffer &
\SynthPercentageAreaPodConvBuffer
\\\cline{2-4} 

&
Input Psum Buffer &
\SynthPercentagePowerPodInputPsumBuffer &
\SynthPercentageAreaPodInputPsumBuffer
\\\cline{2-4} 

&
Output Psum Buffer &
\SynthPercentagePowerPodOutputPsumBuffer &
\SynthPercentageAreaPodOutputPsumBuffer
\\\cline{2-4} 

&
Others &
\SynthPercentagePowerPodOthers &
\SynthPercentageAreaPodOthers

\end{tabular}

}
\caption{Power and area breakdown of the proposed architecture for 256 systolic pods. The design is synthesized in Synopsys Design Compiler using the TSMC 28nm library for up to 16 systolic pods and the results are extrapolated for 256 systolic pods.}
\label{tab:synthesis-results}
\end{table}

\section{Related Work}
\label{sec:related}

Prior work \cite{Hazelwood18, Rhu2016, Wang2018, Yazdanbakhsh2015} proposed software-level support for accelerating DNN workloads exhibiting large degrees of data parallelism on CPUs, GPUs and FPGAs for improved efficiency and latency.
While GPUs equipped with DNN processing support \cite{nvidia_dl2019} are popularly used for training, their long execution pipeline prohibits them from being used as inference accelerators.
Despite the fact that FPGAs are used as DNN accelerators \cite{fowers18, Alwani2016, DiCecco2016, Roukhami2019, peeman13, farabet2011, Wei2017} with high utilization, they do not provide sufficient arithmetic density to achieve high throughput.
Other than existing hardware platforms, custom ASIC solutions that exploit low precision arithmetic and regular memory access patterns have become a popular choice \cite{Albericio2016, Chen2014, Han2016, Reagen2016, Liu20a}.
Many of these ASIC solutions implement certain types of dataflows optimized to improve the power efficiency \cite{kwon18, gao17, gao19, lu17, shao19}.
For example, Eyeriss \cite{chen2016} architecture implements the row-stationary dataflow to reduce the power consumption of data movement.

Many DNN accelerator architectures are based on the variants of systolic arrays \cite{kung82, kung-asplos, baek2020, Choi20, Shomron2019, Liu20a, Liu20b, Drumond2021}.
Google has adopted an ASIC solution and developed Tensor Processing Units (TPU) for its cloud services. TPUv1 is a monolithic systolic array, and it is an inference accelerator \cite{jouppi2017}. TPUv2/v3 target both inference and training, and they consist of two and four systolic arrays sharing a common vector memory \cite{tpu2017}. Although TPUs achieve high throughput, the large systolic arrays used in their designs suffer from underutilization while executing a wide range of DNN workloads, as we explained in Section \ref{sec:optimal-size}. Furthermore, they cannot share a task among a large number of systolic arrays efficiently due to off-chip communication requirements.
Maestro addressed the underutilization problem of the DNN accelerators and proposed a DNN accelerator architecture based on small-sized systolic arrays on 3D interconnect fabric \cite{kung2019, Kung-ISCAS}. However, the proposed architecture suffers from low power efficiency due to reduced computation to memory access ratio.


Prior work has proposed solutions to address various types of underutilization in systolic array-based accelerators. 
AI-MultiTasking \cite{baek2020} and PREMA \cite{Choi20}, both being multi-tenant DNN accelerators, mitigate the underutilization problem that stems from memory stalls and inter-layer dependencies by starting computations, possibly belonging to different tasks, whose dependencies are already satisfied when the computation unit is idle.
SMT-SA \cite{Shomron2019}, a multi-threaded systolic array-based accelerator, addresses the underutilization due to sparsity by using ALUs for non-zero multiplications coming from different matrix multiplication tasks.
Kung et al. \cite{kung-asplos} also address the same issue by proposing a technique to increase the density of sparse weight matrices and a microarchitecture supporting such matrices.
These solutions do not improve the underutilization due to dimension mismatches, and they are orthogonal to the contributions in the paper.



\section{Conclusion}

In this work, we studied multi-pod systolic architectures and their three key design aspects, namely array granularity, interconnect, and tiling. We first analyzed the DNN workloads to reveal their computational requirements and then performed a design space exploration to find the optimal systolic array size. For state-of-the-art DNN workloads from various application domains, we identified that the array size of $32 \times 32$ exhibits the highest effective throughput/Watt among all prior designs. 

Then, we introduced SOSA, which consists of optimally sized systolic arrays. We identified the design requirements for the interconnection between systolic arrays in multi-pod architectures and evaluated various topologies. We showed that Butterfly, which is a multi-stage interconnect, outperforms is the ideal topology for multi-pod accelerators thanks to their scalable hardware cost and short round-trip latency. Then, we proposed a novel tiling scheme to maximize the utilization across multiple pods. We showed that choosing a partitioning size that matches the number of rows in an array improves utilization in a multi-pod system up to $5 \times$ compared to tiling schemes with no partitioning. We demonstrated that the proposed design scales well with the increasing number of pods and achieves up to 600 TeraOps/s at a TDP of 400 Watts with a single-batch DNN workload.

\bibliographystyle{ACM-Reference-Format}
\bibliography{ref}

\appendix

\end{document}